\documentclass[%
 reprint,
 amsmath,amssymb,
 aps,
 pre,
]{revtex4-2}

\usepackage{graphicx}
\usepackage{dcolumn}
\usepackage{bm}

\usepackage{color}

\DeclareTextSymbolDefault{\textquotedbl}{T1}
\providecommand{\tabularnewline}{\\}

\usepackage{xcolor}

\begin{document}
\title{Residual entropy of the dilute Ising chain in a magnetic field}

\author{Yury Panov}
\affiliation{Ural Federal University, 19 Mira street, 620002 Ekaterinburg, Russia}

\date{\today}

\begin{abstract}
The properties of the ground state of the simplest frustrated system, the dilute Ising chain in a magnetic field, are rigorously investigated over the entire range of concentrations of charged non-magnetic impurities. 
Analytical methods are proposed for calculating the residual entropy of frustrated states, including states at phase boundaries, which are based on the Markov property of the system and involve solving a linear optimization problem for energy and a nonlinear optimization problem for entropy. 
These methods allow obvious generalizations for one-dimensional pseudospin models with anisotropic interactions. 
We calculate the composition, entropy and magnetization for the ground state phases. 
We prove the absence of pseudo-transitions in the dilute Ising chain, since the residual entropy of states at phase boundaries is always higher than the entropy of adjacent phases. 
The concentration dependencies of magnetization at the phase boundaries are obtained, and unlike linear dependencies for adjacent phases, they have nonlinear behavior. 
Field-induced transitions between ground states and entropy jumps associated with them are also considered, and in particular, it is shown that the field-induced transition from an antiferromagnetic state to a frustrated one is accompanied by charge ordering.
\end{abstract}


\maketitle

\section{Introduction}

The basis of the unusual behavior of low-dimensional spin and pseudospin systems is the absence or difficulty of long-range order formation. 
This causes the appearance of unique phase states and properties that owe their origin to the special role of fluctuations in these systems. 
On the other hand, the existence of exact solutions, especially for a large number of one-dimensional models, allows us to assess the prospects for obtaining the desired characteristics in simulated physical systems. 
Thus, following the experimental discovery of the striking features of the magnetic behavior of azurite~\cite{Kikuchi2005-prl,Kikuchi2005}, a significant number of works devoted to one-dimensional decorated Ising models appeared. 
These models are characterized by alternating Ising spins and blocks consisting of spins connected by the interaction of Heisenberg type. 
Along with the azurite diamondlike chain~\cite{Honecker2011,Rojas2012-pra,Rojas2012,Ananikian2012,Galisova2013,Torrico2014,Derzhko2015,Richter2015,Lisnyi2015,Torrico2016,Lisnyi2016,Hovhannisyan2016,Torrico2016-prb,Carvalho2018,Carvalho2019,Krokhmalskii2021,Rojas2021}, 
double-tetrahedral chain~\cite{Antonosyan2009,Galisova2015-pre,Galisova2015,DeSouza2018,Galisova2018}, 
ladders~\cite{Strecka2014,Rojas2016,Sousa2018}, 
and tubes~\cite{Alecio2016,Strecka2016} are considered. 
These models demonstrate a lot of fascinating phenomena and reproduce features of real materials, including 
cuprates and vanadates~\cite{Rojas2016}, and 
the heterobimetallic and polymeric coordination compounds~\cite{Strecka2011,Bellucci2014,Torrico2018,Verkholyak2021}. 

One of the intriguing features of decorated Ising chains is the possible presence of frustrated phases in the ground state. 
These phases are similar to the spin ice states in systems with higher dimensions and exhibit various interesting peculiarities of both magnetic response and magnetocaloric properties. 
The possibility of enhancing the magnetocaloric effect in frustrated systems was considered 
in Refs.~\cite{Zhitomirsky2003,Zhitomirsky2004,Sosin2005,Pereira2009}. 

Another striking feature due to the presence of frustrated phases in the ground state of one-dimensional systems are pseudo-transitions. 
They exhibit in the form of a stepwise dependence of entropy on temperature similar to the behavior in phase transitions of the first kind, and a sharp peak in specific heat, which resembles the behavior in phase transitions of the second kind. 
Unlike conventional phase transitions, pseudo-transitions result in an abrupt change in the type of disordered state of a one-dimensional system at a finite temperature, so such thermodynamic characteristics as entropy and specific heat, as well as magnetization and susceptibility, remain continuous functions, although they have very sharp features. 
The universal nature of the pseudo-transition is confirmed by the possibility of defining pseudo-critical exponents~\cite{Rojas2019} having the same values for substantially different systems. 
The pseudo-transition temperature is uniquely determined by the system parameters, such as exchange constants and magnetic field, and this suggests possibilities for both fundamental and practical applications of this phenomenon. 

To predict the existence of a pseudo-transition in a system, it is critically important to know the exact values of entropy in the ground state for all values of the system parameters, and, in particular, at the boundaries between the ground state phases in the phase diagram. 
According to the Rojas rule~\cite{Rojas2020-APP,Rojas2020-BJP}, a pseudo-transition in the system is realized near the phase boundary with a frustrated state if the residual entropy at the phase boundary is equal to the entropy of the frustrated state. 
Such a situation is relatively rare, which causes the narrowness of the range of model parameters for the existence of a pseudo-transition. 

The source of the frustration in magnets, besides geometry, can be impurities. 
The simplest model of such a system is a dilute Ising chain with charged mobile impurities. 
Without taking into account the external magnetic field, the model has an exact solution~\cite{Katsura1965,Rys1969}. 
Its various properties are studied in detail in Refs.~\cite{Kawatra1969,Matsubara1973,Termonia1974}, 
and in the most general form the exact solution is given in Ref.~\cite{Balagurov1974}.
Taking into account the magnetic field, the standard transfer matrix method makes it possible to consider the thermodynamic properties of the model using a numerical solution of a system of nonlinear algebraic equations. 
In this way, the entropy and magnetic Gr\"uneisen parameter of the model were studied at a finite 
temperatures~\cite{Shadrin2022}. 
The properties of the ground state, especially the concentration dependencies, in this case can only be understood at a qualitative level from the numerical solution at low temperatures. 

In the present paper, we propose an analytical method for calculating the residual entropy of a dilute Ising chain in a magnetic field for all possible values of the model parameters, which is based on the Markov property of the model~\cite{Panov2020}. 
Exact analytical expressions for the residual entropy depending on the concentration of impurities are obtained. 
For a given phase of the ground state, the entropy calculation is based on solving a linear optimization problem for the ground state energy. 
For states at phase boundaries, it is necessary to solve an additional nonlinear optimization problem for entropy. 
The proposed method allows to make obvious generalizations for one-dimensional pseudospin models with anisotropic interactions, like the Ising, Potts, Blume-Capel and Blume-Emery-Griffiths models. 
The obtained exact analytical dependencies of the residual entropy allow us to study in detail the nature of the ground state of a dilute Ising chain in a magnetic field, conditions for the existence of pseudo-transitions, and various transitions in the ground state caused by a magnetic field. 
In particular, at certain parameters, a peculiar magnetoelectric effect occurs when a change in the external magnetic field causes a charge ordering of non-magnetic impurities.

The present article is organized as follows. 
In Section~\ref{sec:phase-diagram}, the ground state phase diagrams of the dilute Ising chain with fixed concentration of impurities in external magnetic field are obtained and explored. 
In Section~\ref{sec:entropy}, we present our main results, which are obtained using rigorous methods for calculating the residual entropy, the state compositions, and magnetization for the ground state phases and states at the phase boundaries. 
The transitions induced in the ground state by magnetic field and related effects are considered in Section~\ref{sec:transitions}. 
Finally, conclusions are presented in Section~\ref{sec:conclusion}.

\section{Zero-temperature phase diagram
\label{sec:phase-diagram}}

Phase diagrams at zero temperature of the dilute Ising chain without a magnetic field are presented in Ref.~\cite{Panov2020} in the ''interaction constant''--''chemical potential'' planes. 
Qualitatively, the ground state with accounting for a magnetic field is considered in Ref.~\cite{Shadrin2022}. 
In this Section, we present a rigorous procedure for obtaining of the ground state phase diagrams of the dilute Ising chain with fixed concentration of impurities in external magnetic field. 
Found results will be used in the following Section. 

The Hamiltonian of the model can be written in the following form:
\begin{equation}
	\mathcal{H} = 
	- J \sum_{j=1}^{N} \sigma_{z,j} \sigma_{z,j+1} 
	+ V \sum_{j=1}^{N} P_{0,j} P_{0,j+1} 
	- h \sum_{j=1}^{N} \sigma_{z,j} . 
	\label{eq:H1}
\end{equation}
We use the pseudospin $\sigma=1$ operator, 
where the spin doublet states and impurity correspond 
to the pseudospin $z$-projections $\sigma_z = \pm1$ and $\sigma_z = 0$, respectively,
$J$ is the exchange constant, 
$V>0$ is the effective~\cite{Panov2020} inter-site interaction for impurities,
$P_{0} = 1-\sigma_z^2$ is the projection operator on the impurity state.
We assume that the concentration of non-magnetic charged impurities 
$n = \left\langle \sum_{j} P_{0,j}\right\rangle /N$ is fixed.

For a given $n$, the energy of a dilute Ising chain in a magnetic field can be expressed in terms of the sum over the bonds. 
We introduce $N_{a,b}$ as the number of bonds with the left site in the state $a$ and the right one in the state $b$, so that $\sum_{a,b} N_{a,b} = N$, and determine the concentrations of bonds $x_{a,b}$ by expressions
\begin{equation}
	x_{a,a} = \frac{N_{a,a}}{ N } ,\quad
	x_{a,b} = \frac{N_{a,b} + N_{b,a}}{ N } ,\;
	a \neq b ,
	\label{eq:xab}
\end{equation}
where $\sum_{a,b} x_{a,b} = 1$. 
Here and further, for sums containing $x_{a,b}$, we assume that summation is performed over unordered pairs of indices. 
The functions $x_{a,b}$ depend in general on temperature and all other parameters of the model 
and are expressed in terms of the pair distribution functions for the nearest neighbors. 
The ground state energy per site, $\varepsilon = E/N$, is the linear function of $x_{a,b}$:
\begin{multline}
	\varepsilon = - J \left(x_{1,1}+x_{-1,-1} - x_{1,-1}\right) + V x_{0,0} \\
	- h \left( x_{1,1} - x_{-1,-1} + \tfrac{1}{2}\left(x_{0,1} - x_{0,-1}\right) \right) . 
	\label{eq:e1}
\end{multline}

We introduce the concentration of spin sites $n_s = 1-n = \tfrac{1}{2}-m$, as well as the deviation from half-filling $m$ for the concentration of impurities, $m =n - \frac{1}{2}$.
The concentration of impurities can be expressed in terms of variables $x_{a,b}$ as 
\begin{equation}
	n = x_{0,0} + \tfrac{1}{2}\left(x_{0,1} + x_{0,-1}\right) . 
\end{equation}

Taking into account Eq.~\eqref{eq:xab}, the problem of finding the minimum energy of the ground state takes the canonical form of the linear programming problem:
\begin{equation}
	\left\{
	\begin{array}{l}
		\varepsilon(x_{a,b}) \rightarrow \min , \\
		x_{a,b} \geq 0 , \\
		x_{0,0} + \frac{1}{2}\left(x_{0,1} + x_{0,-1}\right) = n , \\
		x_{1,1} + x_{-1,-1} + x_{1,-1} + \frac{1}{2}\left(x_{0,1} + x_{0,-1}\right) = n_s . 
	\end{array}
	\right.
	\label{eq:emin}
\end{equation}
Here, the energy is the objective function, and solutions of the problem \eqref{eq:emin} correspond to vertices, edges or faces of the feasible polytope of variables $x_{a,b}$.

\begin{table*}[htbp]
\caption{\label{tab:min}
The composition $\left\{x_{a,b}\right\}$ for solutions at vertices 
of the feasible region for the problem~\eqref{eq:emin}. 
The necessary condition for the existence of the solution is shown in the second column, 
and the state energy in the third column.}
\begin{ruledtabular}
\begin{tabular}{ccccccccc}
State & Constraint & $\varepsilon$               & $x_{0,0}$ & $x_{1,1}$ & $x_{-1,-1}$ & $x_{0,-1}$ & $x_{0,1}$ & $x_{1,-1}$ \tabularnewline \hline
1     &            & $-\left(J+h\right) n_s+V n$ & $n$       & $n_s$     & $0$         & $0$        & $0$       & $0$        \tabularnewline
2     &            & $-\left(J-h\right) n_s+V n$ & $n$       & $0$       & $n_s$       & $0$        & $0$       & $0$        \tabularnewline
3     &            & $J n_s+V n$                 & $n$       & $0$       & $0$         & $0$        & $0$       & $n_s$      \tabularnewline
4     & $m<0$      & $2Jm-h n_s$                 & $0$       & $-2m$     & $0$         & $0$        & $2n$      & $0$        \tabularnewline
5     & $m<0$      & $2Jm+h n_s$                 & $0$       & $0$       & $-2m$       & $2n$       & $0$       & $0$        \tabularnewline
6     & $m<0$      & $-2Jm-h n$                  & $0$       & $0$       & $0$         & $0$        & $2n$      & $-2m$      \tabularnewline
7     & $m<0$      & $-2Jm+h n$                  & $0$       & $0$       & $0$         & $2n$       & $0$       & $-2m$      \tabularnewline
8     & $m\geq0$   & $2Vm-h n_s$                 & $2m$      & $0$       & $0$         & $0$        & $2n_s$    & $0$        \tabularnewline
9     & $m\geq0$   & $2Vm+h n_s$                 & $2m$      & $0$       & $0$         & $2n_s$     & $0$       & $0$        \tabularnewline
10    & $m\leq0$   & $2\left(J+h\right) m+h n$   & $0$       & $-2m$     & $0$         & $2n$       & $0$       & $0$        \tabularnewline
11    & $m\leq0$   & $2\left(J-h\right) m-h n$   & $0$       & $0$       & $-2m$       & $0$        & $2n$      & $0$       
\end{tabular}
\end{ruledtabular}
\end{table*}

\begin{figure}
\includegraphics[width=0.42\textwidth]{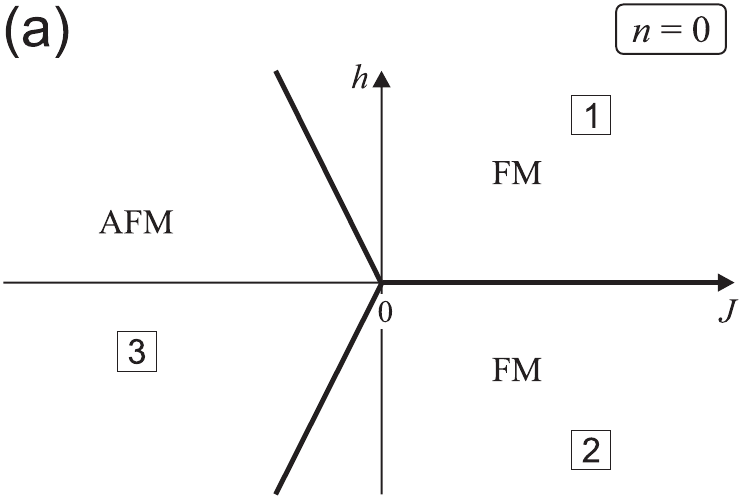}\\[2em]
\includegraphics[width=0.42\textwidth]{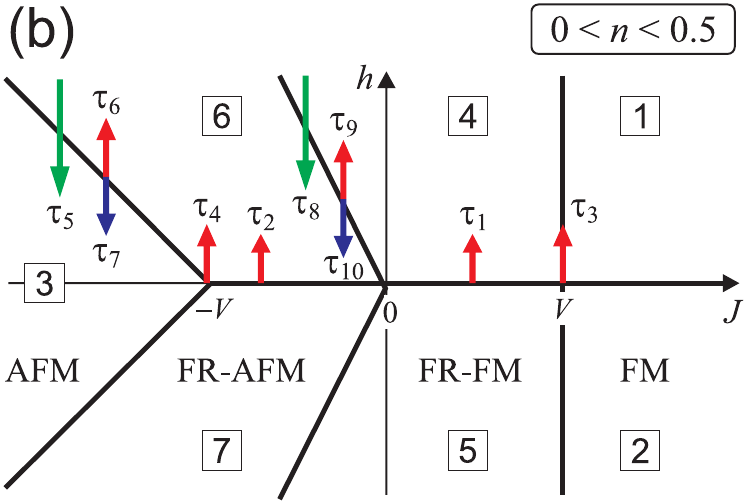}\\[2em]
\includegraphics[width=0.42\textwidth]{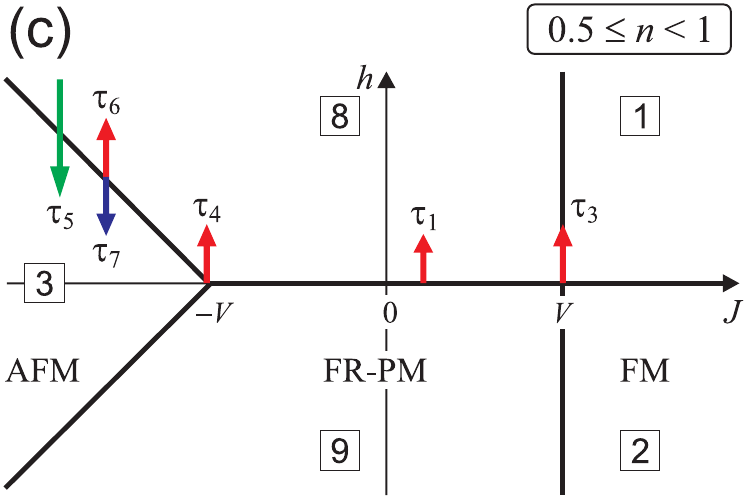}
\caption{
The ground state phase diagrams of a dilute one-dimensional Ising model in a longitudinal magnetic field in the $(h,J)-$ plane for (a) a pure spin chain, $n=0$; (b) the weakly diluted spin chain, $0<n<1/2$; (c) the strongly diluted spin chain, $1/2<n<1$. The framed numbers correspond to the solutions given in Table~\ref{tab:min}. 
The arrows show transitions $\tau_i$ in the ground state which are induced by a change in the external magnetic field. 
The transitions $\tau_1 - \tau_4$ are caused by an increase in the magnetic field from zero to some finite value. 
The transitions $\tau_5$ and $\tau_8$ occur when the magnetic field decreases from values larger to values smaller than the frustration field or the spin-flip field, respectively. 
Transitions $\tau_6$ ($\tau_7$) and $\tau_9$ ($\tau_{10}$) appear when the magnetic field increases (decreases) from the frustration field or the spin-flip field, respectively. 
These transitions are discussed in detail in Section~\ref{sec:transitions}.
}
\label{fig:phdiag}
\end{figure}

The solutions in the vertices of the feasible polytope are listed in Table~\ref{tab:min}. 
They define the regions for the ground state phases in the diagram shown in Fig~\ref{fig:phdiag}. 
The multiplier before $h$ in energy gives the magnetization:
\begin{equation}
	M = x_{1,1} - x_{-1,-1} + \tfrac{1}{2}\left(x_{0,1} - x_{0,-1}\right) . 
	\label{eq:ms}
\end{equation}

Solutions from 1 to 3 exist for all $n$, $0\leq n < 1$. 
In the absence of impurities, at $n=0$, ferromagnetic (FM) ordering (solutions 1 and 2) and antiferromagnetic (AFM) ordering (solution 3) are realized, 
and they are separated by the critical field $|h| = -2J$ (the spin-flip transition field) 
as shown in Fig.~\ref{fig:phdiag}(a). 
The presence of mobile charged impurities qualitatively changes the ground state of the system. 
If $n\neq0$, solutions 1 and 2 describe phases in which macroscopic domains of ferromagnetically ordered spins directed along the field are separated by domains of non-magnetic impurities. 
In this case, $x_{0,0}\neq0$ and $x_{\sigma,\sigma}\neq0$, $\sigma=\pm1$, while $x_{0,\sigma}=0$ in the thermodynamic limit. 
FM phases 1 and 2 have the lowest energy at $J>V>0$, $h\neq0$. 
The magnetization of the FM phases are equal to the concentration of spin sites, $M=n_s$. 
AFM phase 3 is realized if $J<-V-|h|$, and consists of alternating macroscopic domains of antiferromagnetically ordered spins and impurity domains. 
The magnetization of the AFM phase is zero. 

Solutions with numbers from 4 to 7 in Table~\ref{tab:min} exist only for the weakly diluted spin chain, 
$0<n<1/2$, and their energies do not depend on $V$. 
This case is shown in Fig.~\ref{fig:phdiag}(b). 
Solutions 4 and 5 correspond to the minimum energy at $-|h|/2<J<V$, $h>0$ and $h<0$, 
and solutions 6 and 7 have minimal energy at $-V-|h|<J<-|h|/2$, $h>0$ and $h<0$, respectively. 
The equalities $x_{0,0}=0$ and $x_{0,\pm 1}=2n$ indicate that a dilute AFM or FM state is realized, 
where (anti)ferromagnetic clusters of different sizes, including the single spins, are separated by single non-magnetic impurities. 
As will be shown later, these states have nonzero residual entropy, so phases 4 and 5 can be called frustrated ferromagnetic (FR-FM), and phases 6 and 7 are frustrated antiferromagnetic (FR-AFM). 
When the concentration of $n = 1/2$ is reached, a charge-ordered state occurs in which spin and impurity sites alternate. 
For this state, the energy does not depend on the interaction constants, $J$ and $V$. 
Note, that while in FR-FM phases the magnetization equals to the concentration of spin sites, 
$M=n_s$, and decreases with increasing $n$, in the FR-AFM phases $M=n$. 
Concentration dependencies of magnetization are shown in Fig.~\ref{fig:m}. 

Solutions with numbers from 8 and 9 in Table~\ref{tab:min} exist only for the strongly diluted spin chain, $1/2 \leq n<1$, at $-V-|h|<J<V$ (see Fig.~\ref{fig:phdiag}(c)). 
In these states, $x_{\sigma,\sigma'}=0$, $\sigma,\sigma'=\pm1$, and $x_{0,0}=2m$, 
that corresponds to frustrated paramagnetic (FR-PM) phases, where single spins directed along the field are separated by impurity clusters of different sizes, and so $M=n_s$. 
The expressions for the energy of these phases do not contain the exchange interaction constant $J$. 

The energy of solutions 10 and 11 is always higher than the minimum energy at $h\neq0$, but, as will be shown later, 
these solutions are part of the states at the phase boundary $h=0$.

\section{Residual entropy of a dilute Ising chain in a magnetic field
\label{sec:entropy}}

Using the Markov property of the dilute Ising chain~\cite{Panov2020}, 
we write down the probability of the state $(a_1 a_2\ldots a_N)$ 
of a closed chain of $N$ sites ($N\gg1$):
\begin{eqnarray}
	P_{\mathcal{O}}(a_1 a_2 \ldots a_N) &=& P(a_1|a_2)P(a_2|a_3) \ldots P(a_N|a_1) = \nonumber\\
	&=& \prod_{ab} P(a|b)^{N_{ab}} .
	\label{eq:PO}
\end{eqnarray}
Here $P(a|b)$ is the conditional probability that the $i$th site is in the state $a$, provided that the $(i{+}1)$th site is in the state $b$. 
The value of $P(a|b)$ is uniquely related to the sort of bond. 
If $a=b$, we obtain 
\begin{equation}
	x_{a,a} = P(aa) = P(a) P(a|a) 
	\; \Rightarrow \; 
	P(a|a) = \frac{x_{a,a}}{P(a)} . 
\end{equation}
Probabilities of $P(a)$ are equal to the concentrations of the corresponding states and satisfy the equations: 
\begin{equation}
	P(a) = x_{a,a} + \tfrac{1}{2}\sum_{b \neq a} x_{a,b} . 
	\label{eq:Pa}
\end{equation}
Given the equality of the two directions along the chain, we get that for $a\neq b$, 
the equality $N_{a,b}=N_{b,a}=\frac{1}{2}x_{a,b}N$ must be satisfied, 
as well as the equality $P(ab) = P(ba)$, from which follows
\begin{equation}
	x_{a,b} = P(ab) + P(ba) = 2 P(ab) 
	\; \Rightarrow \; 
	P(a|b) = \frac{x_{a,b}}{2P(b)}.
\end{equation}
As a result,
\begin{equation}
	P_{\mathcal{O}} = \prod_{ab} P(a|b)^{N_{ab}} = p_0^N ,
	\label{eq:P0}
\end{equation}
where
\begin{multline}
	p_0 = \left(\frac{x_{0,0}}{P(0)}\right)^{x_{0,0}}
	\left(\frac{x_{1,1}}{P(1)}\right)^{x_{1,1}} 
	\\ \times
	\left(\frac{x_{-1,-1}}{P(-1)}\right)^{x_{-1,-1}} 
	\left(\frac{x_{1,-1}^2}{4P(1)P(-1)}\right)^{x_{1,-1}/2}
	\\ \times
	\left(\frac{x_{0,1}^2}{4P(0)P(1)}\right)^{x_{0,1}/2}
	\left(\frac{x_{0,-1}^2}{4P(0)P(-1)}\right)^{x_{0,-1}/2} . 
	\label{eq:p0}
\end{multline}

Equation~(\ref{eq:P0}) is valid for any temperature, 
but at zero temperature it also provides a way to explicitly calculate entropy. 

The ground state energy is completely determined by the values of $x_{a,b}$, according to Eq.~\eqref{eq:e1}. 
We will assume that the microcanonical distribution is valid for the ground state, that is, all states with a given energy have equal probabilities. 
The sum of these probabilities is $1$, that makes it possible to find the statistical weight $\Gamma$ of the ground state:
\begin{equation}
	P(E(x_{a,b})) = 1 = \Gamma \, P_{\mathcal{O}} ,
\end{equation}
and residual entropy:
\begin{equation}
	s_0 = \frac{\ln\Gamma}{N} = - \ln p_0 .
\end{equation}
Taking into account Eq.~\eqref{eq:p0}, we obtain:
\begin{equation}
	s_0 = - \sum_{a,b} x_{a,b} \ln x_{a,b} + P_2 \ln2 + \sum_a P(a) \ln P(a) , 
	\label{eq:s0-common}
\end{equation}
where probabilities $P(a)$ are defined by Eq.~\eqref{eq:Pa}, in particular, $P(0)=n$, 
and the total concentration $P_2$ of pairs of different states is introduced: 
\begin{equation}
	P_2 = x_{1,-1}+x_{0,1}+x_{0,-1} . 
\end{equation}

Equation~\eqref{eq:s0-common} allows us to find the concentration dependence of the residual entropy with a known composition $\left\{x_{a,b}\right\}$ for the ground state. 
To solve this problem within the framework of the standard approach, it is necessary to find the largest eigenvalue of the transfer matrix, determine the parametric dependence of entropy on concentration using the chemical potential as a parameter, and find the limit at zero temperature. 
For a dilute Ising chain in a magnetic field, this can only be done numerically~\cite{Shadrin2022}, while Eq.~\eqref{eq:s0-common} provides an exact analytical result. 

Using the solutions in Table~\ref{tab:min}, we obtain expressions for the residual entropy of phases 1-9. 
The FM and AFM solutions 1-3 have zero entropy. 
Solutions from 4 to 9 have nonzero residual entropy for all impurity concentrations, 
except for the marginal ones, $n=0$, $\tfrac{1}{2}$ and $1$. 
The entropy of the FR-FM (numbers 4 and 5) and FR-PM (numbers 8 and 9) solutions has the same dependency on $|m|$, demonstrating a kind of symmetry of impurity and spin states in the FM case:
\begin{multline}
	s_0 =
	- 2|m| \ln\left(2|m|\right)
	- \left(\tfrac{1}{2}-|m|\right) \ln\left(\tfrac{1}{2}-|m|\right) \\
	+ \left(\tfrac{1}{2}+|m|\right) \ln\left(\tfrac{1}{2}+|m|\right) . 
	\label{eq:s0-4589}
\end{multline}
For a given concentration, this value is greater than entropy of the FR-AFM solutions (numbers 6 and 7):
\begin{equation}
	s_0 = 
	- |m|\ln|m| 
	- \left(\tfrac{1}{2}-|m|\right)\ln\left(\tfrac{1}{2}-|m|\right)
	- \tfrac{1}{2}\ln2 . 
	\label{eq:s0-67}
\end{equation}

\begin{figure*}[htbp]
\includegraphics[width=0.29\textwidth]{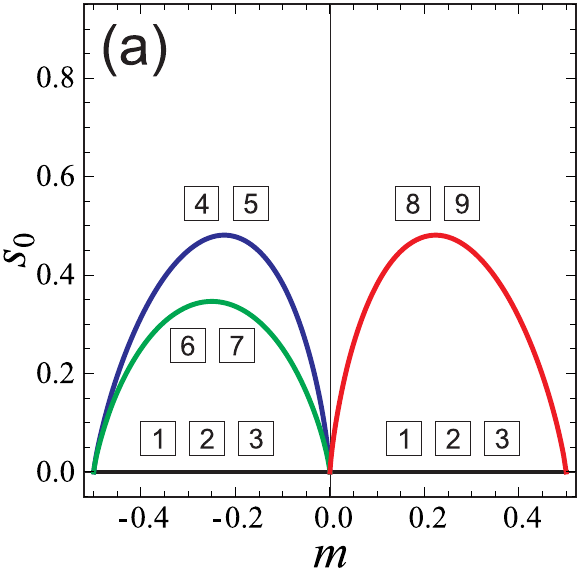} \quad
\includegraphics[width=0.29\textwidth]{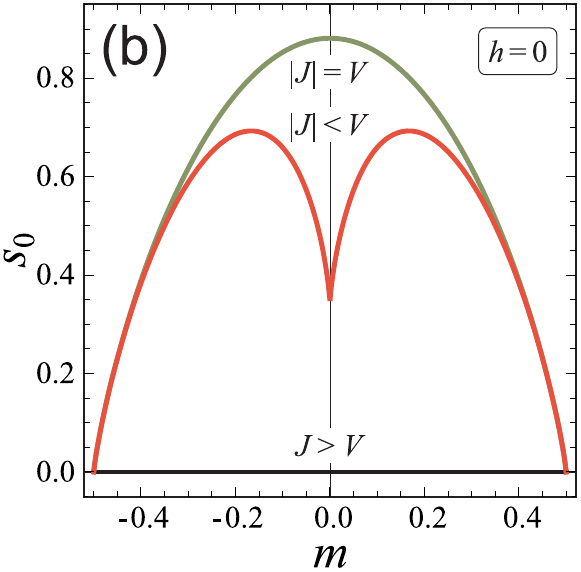} \quad
\includegraphics[width=0.29\textwidth]{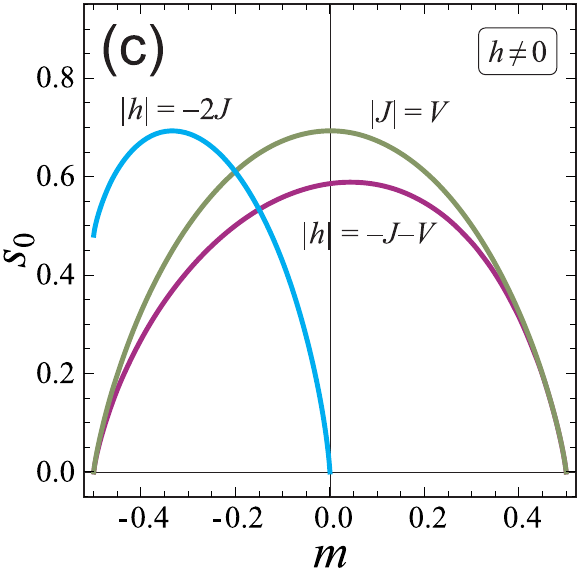} 
\caption{
Concentration dependencies of the residual entropy of a dilute Ising chain for 
(a) the states inside the regions of the ground state phases 
(the framed numbers correspond to the solutions given in Table~\ref{tab:min}), 
(b) the states at the phase boundary at $h=0$ given in Table~\ref{tab:sol0h}, 
(c) the states at the phase boundaries at $h\neq0$ given in Table~\ref{tab:solbh}.
Next to the curves are the equations of the corresponding phase boundaries in the ground state diagram.
\label{fig:s0}
}
\end{figure*}

The dependencies of the residual entropy on $m$ for solutions 1 to 9 are shown in Fig.~\ref{fig:s0}(a). 
The obtained dependencies are consistent with the results for entropy at low temperatures, which were obtained by numerically solving a system of nonlinear algebraic equations within the framework of a grand canonical ensemble~\cite{Shadrin2022}. 
Within the framework of the method presented here, it is possible to explore the behavior of functions analytically. 
The function~\eqref{eq:s0-4589} has maxima 
$s_{0,max} = -\frac{1}{2} \ln \frac{\sqrt{5}-1}{\sqrt{5}+1} \approx 0.481$ 
at $m = \pm\frac{1}{2\sqrt{5}} \approx \pm0.224$, 
and the function~\eqref{eq:s0-67} has a maximum 
$s_{0,max} = \frac{1}{2} \ln 2 \approx 0.347 $ 
at $m = -\frac{1}{4}$.

At the phase boundary, the energies of adjacent phases are equal, so the boundary state should be a superposition of these phases, unless this leads to an increase in energy. 
We define coefficients $c_{\nu}$ to be the variational parameters in linear combinations 
$x_{a,b} = \sum_{\nu} c_{\nu} x_{a,b}^{(\nu)}$, 
where $x_{a,b}$ are unknown concentrations of the boundary state, 
and $x_{a,b}^{(\nu)}$ are the found solutions for adjacent phases.
The coefficients $c_{\nu}$ are determined from the principle of maximum entropy. 
Using Eq.\eqref{eq:s0-common} for $s_0$, we obtain a nonlinear optimization problem:
\begin{equation}
	\left\{
	\begin{array}{l}
		s_0(c_{\nu}) \rightarrow \max , \\
		c_{\nu} \geq 0 , \\
		\sum_{\nu} c_{\nu} = 1 . 
	\end{array}
	\right.
	\label{eq:S0max}
\end{equation}

\begin{table*}[htbp]
\caption{\label{tab:sol0h}
The composition $\left\{x_{a,b}\right\}$ of the ground states at phase boundary at $h=0$. 
The value $x^{\ast}$ is defined by Eq.\eqref{eq:xast}. 
The necessary condition for the existence of a state is shown in the first column, 
and the numbers of contributing phases in the second column.}
\begin{ruledtabular}
\begin{tabular}{cccccccc}
Constraint & Phases & $x_{0,0}$ & $x_{1,1}$ & $x_{-1,-1}$ & $x_{0,-1}$ & $x_{0,1}$ & $x_{1,-1}$ \tabularnewline \hline
$J>V$ & 1, 2 & $n$ & $\tfrac{1}{2} n_s$ & $\tfrac{1}{2} n_s$ & $0$ & $0$ & $0$ \tabularnewline \hline
$0<J<V$, $m<0$ & $4, 5, 10,$ 11 & $0$ & $-m$ & $-m$ & $n$ & $n$ & $0$ \tabularnewline
$-V<J<0$, $m<0$ & 6, 7 & $0$ & $0$ & $0$ & $n$ & $n$ & $-2m$ \tabularnewline
$|J|<V$, $m>0$ & 8, 9 & $2m$ & $0$ & $0$ & $n_s$ & $n_s$ & $0$ \tabularnewline \hline
$J=V$, $m<0$ & 1, 2, 4, $5, 10, 11$ & $2m+x^{\ast}$ & $\tfrac{1}{2}x^{\ast}$ & $\tfrac{1}{2}x^{\ast}$ & $n_s-x^{\ast}$ & $n_s-x^{\ast}$ & $0$ \tabularnewline
$J=V$, $m>0$ & 1, 2, 8, 9 & $2m+x^{\ast}$ & $\tfrac{1}{2}x^{\ast}$ & $\tfrac{1}{2}x^{\ast}$ & $n_s-x^{\ast}$ & $n_s-x^{\ast}$ & $0$ \tabularnewline
$J=-V$, $m<0$ & 3, 6, 7 & $2m+x^{\ast}$ & $0$ & $0$ & $n_s-x^{\ast}$ & $n_s-x^{\ast}$ & $x^{\ast}$ \tabularnewline
$J=-V$, $m>0$ & 3, 8, 9 & $2m+x^{\ast}$ & $0$ & $0$ & $n_s-x^{\ast}$ & $n_s-x^{\ast}$ & $x^{\ast}$
\end{tabular}
\end{ruledtabular}
\end{table*}

Results for the boundary line $h=0$ are listed in Table~\ref{tab:sol0h}. 
One can see that the composition of the states for the weakly diluted spin chain,
$m<0$, at $0<J\leq V$, includes solutions 10 and 11 from Table~\ref{tab:min}. 
The parameter 
\begin{equation}
	x^{\ast} = \sqrt{\tfrac{1}{2} + 2m^2} - \tfrac{1}{2} - m 
	\label{eq:xast}
\end{equation}
is equal to the concentration of antiferromagnetically ordered spin pairs at $J=-V$ 
and the concentration of ferromagnetically ordered spin pairs at $J=V$.

Solutions in Table~\ref{tab:sol0h} are divided into 3 groups. 
It is interesting to note that for significantly different compositions, 
the entropy within the group is defined by identical dependencies on $m$. 
The FM states at $J>V$ have zero entropy. 
The entropy of states at the interval $|J|<V$ has the following form:
\begin{multline}
	s_0 =
	- 2|m| \ln\left(2|m|\right)
	- \left(\tfrac{1}{2}-|m|\right) \ln\left(\tfrac{1}{2}-|m|\right) \\
	+ \left(\tfrac{1}{2}+|m|\right) \ln\left(\tfrac{1}{2}+|m|\right) 
	+ \left(\tfrac{1}{2}-|m|\right) \ln2 . 
	\label{eq:s0h0-1}
\end{multline}
Both FM and AFM states in the points $J=\pm V$, $h=0$, have the same entropy:
\begin{equation}
	s_0 = \left(\tfrac{1}{2}+m\right) \ln \frac{\tfrac{1}{2}+m}{2m+x^{\ast}}
	+\left(\tfrac{1}{2}-m\right) \ln \frac{\tfrac{1}{2}-m}{x^{\ast}} . 
	\label{eq:s0h0-2}
\end{equation}

Fig.~\ref{fig:s0}(b) shows the dependencies~\eqref{eq:s0h0-1} and~\eqref{eq:s0h0-2}. 
The function~\eqref{eq:s0h0-1} has two maxima 
$s_{0,max} = \ln 2  \approx 0.693$ at $m = \pm\frac{1}{6}$ 
and the local minimum $s_{0,max} = \frac{1}{2} \ln 2 \approx 0.347$ at $m = 0$, 
the function~\eqref{eq:s0h0-2} has maximum 
$s_{0,max} = \ln \left(1+\sqrt{2}\right) \approx 0.881 $ at $m = 0$. 

The concentration dependencies of entropy \eqref{eq:s0h0-1} and \eqref{eq:s0h0-2} 
coincide with those obtained earlier~\cite{Panov2020} 
from the exact solution for a dilute Ising chain in the zero field as the limit at $T\rightarrow0$.
This confirms the correctness of the general equation for the residual entropy~\eqref{eq:s0-common} 
and the method of obtaining entropy for the boundary states~\eqref{eq:S0max}.

Solutions at the boundaries between the ground state phases at $h\neq0$ 
are listed in Table~\ref{tab:solbh}. 
Here $\alpha$ fulfills the equation
\begin{equation}
	\left(1 - \mu \alpha\right)\sqrt{1 - \alpha^2} = 2 \mu \alpha^2 , \qquad
	\mu = \frac{1-2m}{1+2m} . 
	\label{eq:alpha-eq}
\end{equation}
If $m>0$, then $0\leq\alpha\leq1$, 
and if $m<0$, then $0\leq\alpha\leq1/\mu$. 
The parameters 
\begin{eqnarray}
	x_0 &=& \left(\tfrac{1}{2}+m\right)\left(1-\mu\alpha\right) , \\ 
	x_1 &=& \left(1-2m\right) \alpha , \\ 
	x_2 &=& \left(\tfrac{1}{2}-m\right)\left(1-\alpha\right) , 
	\label{eq:x123}
\end{eqnarray}
are equal to concentrations of the impurity pairs, impurity-spin pairs, and
antiferromagnetically ordered spin pairs, respectively, at the phase boundary $J=-V-|h|$. 
The concentration of antiferromagnetically ordered spin pairs at the spin-flip boundary 
is also introduced:
\begin{equation}
	x^{\ast\ast} = \tfrac{1}{5} \left(\tfrac{1}{2} - 9m - \sqrt{\tfrac{1}{4}-9m+m^2}\right) . 
	\label{eq:xastast}
\end{equation}

\begin{table*}[htbp]
\caption{\label{tab:solbh}
The composition $\left\{x_{a,b}\right\}$ of the ground state at phase boundaries at $h\neq0$. 
The values $x_0$, $x_1$, $x_2$, and $x^{\ast\ast}$ are defined by Eq.~\eqref{eq:x123} 
and~\eqref{eq:xastast}. 
The necessary condition for the existence of a state is shown in the first column, 
and the numbers of contributing phases in the second column.}
\begin{ruledtabular}
\begin{tabular}{cccccccc}
Constraint      & Phases & $x_{0,0}$ & $x_{1,1}$          & $x_{-1,-1}$        & $x_{0,-1}$ & $x_{0,1}$ & $x_{1,-1}$     \tabularnewline \hline
$J=V$           &        &           &                    &                    &            &           &                \tabularnewline
$m<0$, $h>0$    & 1, 4   & $n^2$     & $n_s^2$            & $0$                & $0$        & $2nn_s$   & $0$            \tabularnewline
$m<0$, $h<0$    & 2, 5   & $n^2$     & $0$                & $n_s^2$            & $2nn_s$    & $0$       & $0$            \tabularnewline
$m\geq0$, $h>0$ & 1, 8   & $n^2$     & $n_s^2$            & $0$                & $0$        & $2nn_s$   & $0$            \tabularnewline
$m\geq0$, $h<0$ & 2, 9   & $n^2$     & $0$                & $n_s^2$            & $2nn_s$    & $0$       & $0$            \tabularnewline \hline
$J=-V-|h|$      &        &           &                    &                    &            &           &                \tabularnewline
$m<0$, $h>0$    & 3, 6   & $x_0$     & $0$                & $0$                & $0$        & $x_1$     & $x_2$          \tabularnewline
$m<0$, $h<0$    & 3, 7   & $x_0$     & $0$                & $0$                & $x_1$      & $0$       & $x_2$          \tabularnewline
$m\geq0$, $h>0$ & 3, 8   & $x_0$     & $0$                & $0$                & $0$        & $x_1$     & $x_2$          \tabularnewline
$m\geq0$, $h<0$ & 3, 9   & $x_0$     & $0$                & $0$                & $x_1$      & $0$       & $x_2$          \tabularnewline \hline
$J=-|h|/2$      &        &           &                    &                    &            &           &                \tabularnewline
$m<0$, $h>0$    & 4, 6   & $0$       & $-2m-x^{\ast\ast}$ & $0$                & $0$        & $2n_s$    & $x^{\ast\ast}$ \tabularnewline
$m<0$, $h<0$    & 5, 7   & $0$       & $0$                & $-2m-x^{\ast\ast}$ & $2n_s$     & $0$       & $x^{\ast\ast}$
\end{tabular}
\end{ruledtabular}
\end{table*}

The states at the boundary between FM and frustrated phases, $J=V$, $h\neq0$, have the entropy 
\begin{equation}
	s_0 = - \left(\tfrac{1}{2} - m\right) \ln \left(\tfrac{1}{2} - m\right) 
	- \left(\tfrac{1}{2} + m\right) \ln \left(\tfrac{1}{2} + m\right) . 
	\label{eq:s0:J=V}
\end{equation}
This function is symmetric with respect to a line $m = 0$ 
and has a maximum $s_{0,max} = \ln 2  \approx 0.693$ at $m = 0$. 

The field $|h|=-J-V$ (where $J<-V<0$) can be called the frustration field, 
since this field defines the boundary between the AFM and frustrated phases.
The entropy at this boundary has the following form:
\begin{equation}
	s_0 = -\left(\tfrac{1}{2}+m\right) \ln \left(1-\mu\alpha\right)
	+\tfrac{1}{2}\left(\tfrac{1}{2}-m\right) \ln \frac{1+\alpha}{1-\alpha} . 
	\label{eq:s0:J=-V-|h|}
\end{equation}
This function has no symmetry with respect to a line $m = 0$ 
and reaches a maximum $s_{0,max} \approx 0.589$ at $m = 0.043$. 

At the spin-flip boundary, $m<0$, $J=-|h|/2$, $h\neq0$, the entropy is given by
\begin{multline}
	s_0 = - \left(\tfrac{1}{2}+m\right) \ln \left(1+2m\right) \\
	+ \tfrac{1}{2} \ln \left(1-2m-x^{\ast\ast}\right)
	+ m \ln x^{\ast\ast} . 
	\label{eq:s0:J=-|h|/2}
\end{multline}
In this case, the maximum $s_{0,max} = \ln 2  \approx 0.693$ is attained at $m = -\frac{1}{3}$. 

The concentration dependencies~(\ref{eq:s0:J=V}--\ref{eq:s0:J=-|h|/2}) 
are shown in Fig.~\ref{fig:s0}(c).

In all the cases considered, the entropy of states at the boundary between the ground state phases is higher than the entropy of the adjacent phases. 
Using the Rojas rule~\cite{Rojas2020-APP,Rojas2020-BJP}, we can conclude that there are no pseudo-transitions in the one-dimensional dilute Ising model.

The magnetization at the boundaries between the phases of the ground state can be found from the Eq.~\eqref{eq:ms} using the solutions in Tables~\ref{tab:sol0h} and \ref{tab:solbh}. 
All solutions in Table~\ref{tab:sol0h} have zero magnetization. 
The magnetization at the boundary between FM and FR-FM phases, $J=V$, $h\neq0$, coincides with that for these phases, $M = n_s$.
At the frustration field boundary, $J=-V-|h|$, $h\neq0$, 
we obtain $M = \alpha n_s$, 
where $\alpha$ is defined by Eq.~\eqref{eq:alpha-eq}. 
At the spin-flip boundary, $m<0$, $J=-|h|/2$, $h\neq0$, the magnetization has the following form:
\begin{equation}
	M = n_s - x^{\ast\ast} 
	= \tfrac{1}{5} \left( 2 + 4m + \sqrt{\tfrac{1}{4}-9m+m^2} \right) . 
	\label{eq:ms-spin-flip}
\end{equation}

\begin{figure}[htbp]
\includegraphics[width=0.33\textwidth]{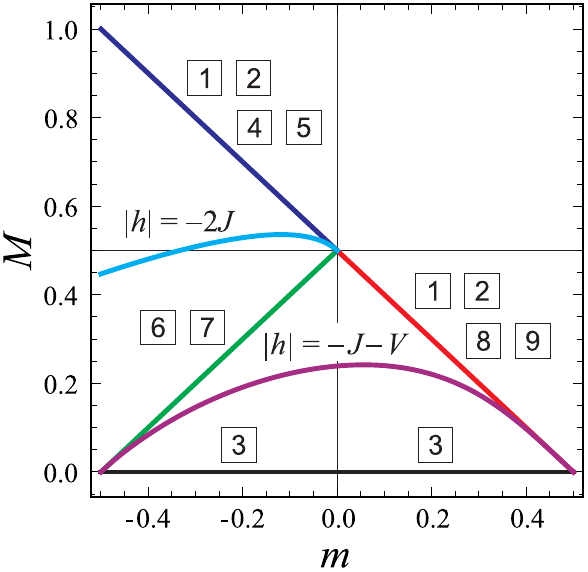} 
\caption{Concentration dependencies of magnetization of a dilute Ising chain. 
Magnetization for the states inside the regions of the ground state phases is described by linear dependencies 
(the framed numbers correspond to the solutions given in Table~\ref{tab:min}). 
Magnetization at the phase boundaries demonstrates nonlinear concentration dependencies 
(the equations of the phase boundaries in the ground state diagram are given near the curves). 
\label{fig:m}} 
\end{figure}

Fig.~\ref{fig:m} shows the concentration dependencies of magnetization 
for the phase states and boundary states at $h\neq0$. 
At the boundaries, the magnetization demonstrates a nonlinear concentration dependence and has an intermediate value relative to the magnetization of adjacent phases. 
The spin-flip boundary magnetization~\eqref{eq:ms-spin-flip}  
for the pure Ising chain equals 
$M_{0} = \frac{1}{\sqrt{5}} \approx 0.447$ and attains maximum 
$M_{max} = 4-2\sqrt{3} \approx 0.536$ at $m = \frac{9}{2} - \frac{8}{\sqrt{3}} \approx - 0.119$. 
At $m=0$, phases FR-AFM and FR-FM transform into the FR-PM phase, so that all 3 dependencies merge into one, $M = n_s$. 
At the boundary between AFM and frustrated phases, $J=-V-|h|$, $h\neq0$, 
the magnetization curve is not symmetric with respect to a line $m = 0$ and has the maximum $M_{max} \approx 0.242$ at $m \approx 0.055$.

\section{Transitions between the ground state phases induced by a magnetic field 
\label{sec:transitions}}

In this section, we study transitions between the ground state phases, 
which can be caused by a change in the magnetic field, 
and in particular, the jump in entropy in these transitions, 
$\Delta s_0 = s_0(initial\;state) - s_0(final\;state)$, 
which gives information about the magnetocaloric properties of the system. 

\begin{figure*}[htbp]
\includegraphics[width=0.29\textwidth]{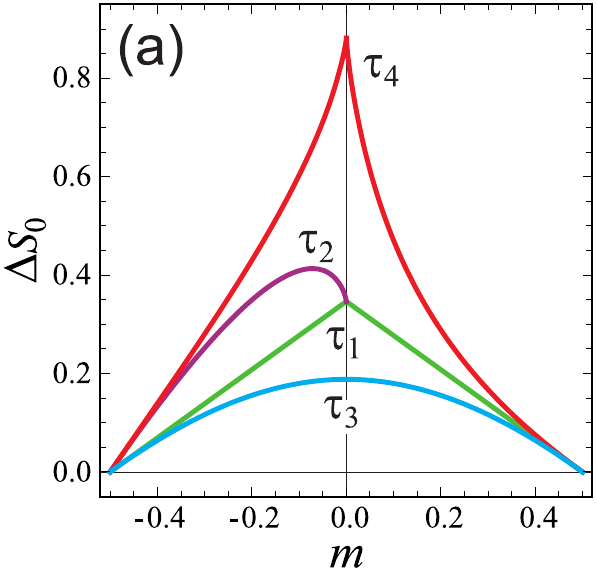} \qquad 
\includegraphics[width=0.29\textwidth]{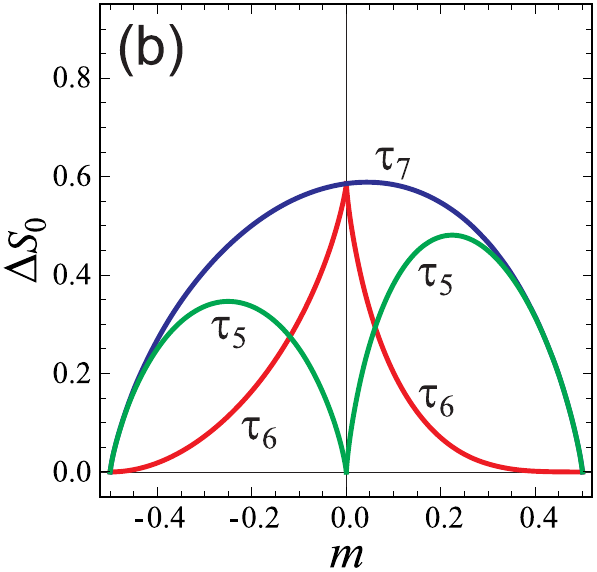} \qquad 
\includegraphics[width=0.29\textwidth]{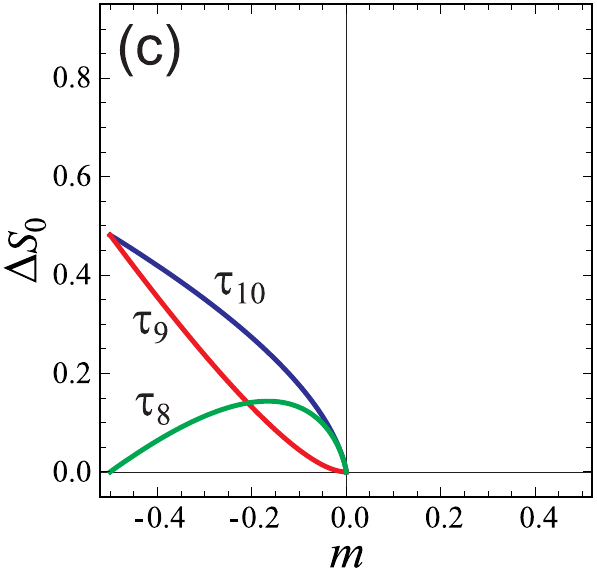} 
\caption{Residual entropy jumps induced by a magnetic field for transitions 
(a) $\tau_1-\tau_4$ near zero field, 
(b) $\tau_5-\tau_7$ near the frustration field, 
(c) $\tau_8-\tau_{10}$ near the spin-flip field. 
The transitions are shown by arrows in Fig.~\ref{fig:phdiag}.
\label{fig:tr}
}
\end{figure*}

Fig.~\ref{fig:phdiag} shows different field-induced transitions in the ground state, 
which can be divided into 3 groups. 
The first group consists of transitions $\tau_1 - \tau_4$ from the states at the phase boundary 
$-V \leq J \leq V$, $h=0$, into frustrated states at $h \neq 0$. 
In the strongly diluted case, $1/2 \leq n < 1$, the final state is the same for both $J>0$ and $J<0$, 
so only the transition $\tau_1$ remains. 
Fig.~\ref{fig:s0}(a,b) shows that the entropy of the system in the field is always lower than without the field. 
The entropy jumps $\Delta s_0 = s_0(h=0) - s_0(h\neq0)$ for the transitions $\tau_1 - \tau_4$ are shown in Fig.~\ref{fig:tr}(a). 
The value $\Delta s_0$ has maximum for the FR-FM and FR-PM phases (the transition $\tau_1$) 
at a half-filling, $m=0$, and for the FR-AFM (the transition $\tau_2$) at some $m<0$. 
The maximum jump in the residual entropy is achieved in the transition $\tau_4$ at $m=0$.
This happens due to the transition from the state at $-J=V$, which is completely frustrated due to the compensation in energy of contributions from the exchange and charge interactions, to the charge ordered state, which induced by the magnetic field at a half-filling. 
The value $\Delta s_0$ in this case is significantly higher than for FR-FM and FR-PM phases in the transition $\tau_1$, since at $|J|<V$, $m=0$ and $h=0$ the ground state is also partially charge-ordered.
The $\Delta s_0$ have small value in the transition $\tau_3$, because
the nonzero magnetic field leaves the state frustrated for all $m$ at $J=V>0$ , 
except for the values of $m=\pm1/2$.

Fig.~\ref{fig:tr}(b) shows the concentration dependencies of the residual entropy jumps 
near the frustration field boundary, in transitions $\tau_5 - \tau_7$. 
Since for the AFM phase $s_0=0$, 
$\Delta s_0$ for the transitions $\tau_5$ and $\tau_7$ coincide with the residual entropy of the corresponding frustrated phases, 
and for $\Delta s_0$ the obvious equality holds: 
$\Delta s_0( \tau_7 ) = \Delta s_0( \tau_6 ) + \Delta s_0( \tau_5 )$. 

The transition, the reverse $\tau_5$, from AFM to FR-AFM or FR-PM phase demonstrates a kind of magnetoelectric effect: when the value of the magnetic field increases more than the value of the frustration field, $|h|=-J-V$, a charge ordering appears in the system. 
The markers of the charge ordering are the nonzero values $x_{0,\pm1}$ in the FR-AFM and FR-PM phases (see Table~\ref{tab:min}), while in the AFM phase $x_{0,\pm1}=0$. 
The charge order parameter reaches maximum at half filling, $m=0$, and in this case the change of the ground state will manifest itself most distinctly: the dilute AFM state at $|h|<-J-V$, which consists of macroscopic AFM and impurity domains and has zero magnetization, is replaced by a charge-ordered state at $|h|>-J-V$, in which the spin and impurity sites alternate and the magnetization is $M=1/2$.

Fig.~\ref{fig:tr}(c) shows the concentration dependencies of $\Delta s_0$ for transitions $\tau_8$, $\tau_9$, and $\tau_{10}$ near the spin-flip boundary. 
In this case, $\Delta s_0$ has maximum at $m=-1/2$, i.e. in the absence of impurities.
For $-1/2<m<0$, the value $\Delta s_0$ is greater when switching to the FR-AFM phase than when switching to the FR-FM phase, and for $\Delta s_0$, for each value of $m$, the equality $\Delta s_0( \tau_{10} ) = \Delta s_0( \tau_9 ) + \Delta s_0( \tau_8 )$ is satisfied.
If $-1/2<m<0$, the entropy jump is greater for the transition to the FR-AFM phase than for the transition to the FR-FM phase. 
For all $m$, the equality $\Delta s_0( \tau_{10} ) = \Delta s_0( \tau_9 ) + \Delta s_0( \tau_8 )$ holds.

\begin{table}[bhtp]
\caption{\label{tab:maxDs}
Critical concentrations $m_0$ and the maximum jumps of entropy $\Delta s_{max}$ for the field-induced transitions shown in Fig.~\ref{fig:phdiag}.}
\begin{ruledtabular}
	\begin{tabular}{ c c c l }
	 Transition & $m_0$ & $\Delta s_{max}$ & \tabularnewline  \hline
	$\tau_1$ & $0$ & $\tfrac{1}{2} \ln 2$ & $\approx 0.347$ \tabularnewline[0.25em] 
	$\tau_2$ & $-\tfrac{1}{14}\approx -0.071$ & $\ln \tfrac{4}{\sqrt{7}}$ & $\approx 0.413$ \tabularnewline[0.5em] 
	$\tau_3$ & $0$ & $\ln \tfrac{\sqrt{2}+1}{2}$ & $\approx 0.188$ \tabularnewline[0.25em] 
	$\tau_4$ & $0$ & $\ln \left(\sqrt{2}+1\right)$ & $\approx 0.881$ \tabularnewline[0.25em] 
	$\tau_5$ & $-\tfrac{1}{4}$ & $\tfrac{1}{2} \ln 2$ & $\approx 0.347$ \tabularnewline[0.25em] 
	 $\tau_5$ & $\tfrac{1}{2\sqrt{5}}\approx 0.224$ & $\tfrac{1}{2} \ln \tfrac{\sqrt{5}+1}{\sqrt{5}-1}$ & $\approx 0.481$ \tabularnewline[0.25em] 
	$\tau_6$ & $0$ &  & $\approx 0.586$ \tabularnewline[0.25em] 
	$\tau_7$ & $0.043$ &  & $\approx 0.588$ \tabularnewline[0.25em] 
	$\tau_8$ & $-\tfrac{1}{6}$ & $\ln \tfrac{2}{\sqrt{3}}$ & $\approx 0.144$ \tabularnewline[0.25em] 
	$\tau_9$, $\tau_{10}$ & $-\tfrac{1}{2}$ & $\tfrac{1}{2} \ln \tfrac{\sqrt{5}+1}{\sqrt{5}-1}$ & $\approx 0.481$ \tabularnewline[0.25em] 
\end{tabular}
\end{ruledtabular}
\end{table}

Table~\ref{tab:maxDs} shows the values of the critical concentration $m_0$ and maximal jumps of entropy $\Delta s_{max}$ 
for dependencies shown in Fig.~\ref{fig:tr}.
The maximum value of $\Delta s_{max} \approx 0.881$ can be reached, 
when the magnetic field is turned on for the half filled with non-magnetic impurities chain 
in the AFM frustration point, $J=-V$.
The second largest value of $\Delta s_{max} \approx 0.588$ is realized for a strongly diluted AFM phase, 
$J<-V$ and $n=0.543$, when the frustration field $|h|=-J-V$ is reached.
This value slightly exceeds the entropy jump $\Delta s_{max}\approx.586$ at half filling, 
which is the same both with increasing and decreasing the field counted from the frustration field.
The maximum value of $\Delta s_{max} \approx 0.481$ in the absence of impurities 
is achieved in the AFM chain at the spin-flip field $|h|=-2 J$, 
which is very large for typical values of the exchange constant.

\section{Conclusion
\label{sec:conclusion}}

The dilute Ising chain is the simplest model of systems frustrated due to impurities.
Such systems are of interest from both fundamental and applied points of view. 
Although the exact solution of this model without a magnetic field has been studied in detail earlier~\cite{Katsura1965,Rys1969,Kawatra1969,Matsubara1973,Termonia1974,Balagurov1974}, and the thermodynamic properties taking into account the magnetic field can be calculated by the transfer matrix method~\cite{Shadrin2022}, some subtle details of the structure of the ground state phase diagram are difficult to calculate in the conventional approach, but they are important for predicting possible features of thermodynamic behavior. 
These include the properties of states at the boundaries between the ground state phases in the phase diagram. 

Here we presented the calculation of the zero-temperature phase diagram of a dilute Ising chain in a magnetic field at a fixed impurities concentration as a solution of the linear optimization problem, as well as an equation for the entropy of frustrated states based on the Markov property of the system, and a method for calculating the entropy of states at phase boundaries as a solution of the nonlinear optimization problem. 
The methods allow generalization to other one-dimensional models with Ising-type interactions. 
The zero-temperature phase diagram was obtained and the composition, magnetization and entropy of the ground state phases were investigated for all possible values of the system parameters. 
If $h\neq0$, the ground state of the system is frustrated at $-|h|-V<J<V$ and $n\neq 0, 0.5, 1$, while if $J>V$ or $J<-|h|-V$, the residual entropy is zero. 
The concentration dependencies of the residual entropy of frustrated phases are consistent with the numerical results for entropy at low temperatures~\cite{Shadrin2022}, and at the phase boundaries $h=0$, the exact analytical results obtained earlier in Ref.~\cite{Panov2020} are reproduced, which is the test of the methods used. 
It was found that the residual entropy of states at phase boundaries is always higher than the entropy of adjacent phases, which, according to the Rojas rule~\cite{Rojas2020-APP,Rojas2020-BJP}, means the absence of pseudo-transitions in the dilute Ising chain.  
Microscopically, this is due to the absence of phases for which mixing at the phase boundary is prohibited due to an increase in energy~\cite{Panov2021}. 
For the states at the phase boundaries, the exact analytical dependencies of magnetization on the impurities concentration were investigated. 
They exhibit nonlinear behavior, although for adjacent phases the magnetization is linear in concentration. 
Transitions between the ground states induced by changes in the external magnetic field were considered. 
It was found that when passing through the phase boundary determined by the frustration field $|h|=-J-V$, the charge ordering induced by magnetic field occurs in the system. 
The maximum of this effect is observed when the chain is half filled with impurities. 
The entropy jump in transitions induced by a magnetic field is also considered. 
This value reaches a maximum, when the magnetic field is turned on for the half filled AFM chain 
at $J=-V$. 
Comparison of the spin-flip field $|h|=-2 J$, with the frustration field $|h|=-J-V$, shows the advantage of AFM systems diluted with mobile charged non-magnetic impurities to obtain the maximum jump in entropy.



\begin{thebibliography}{49}%
\makeatletter
\providecommand \@ifxundefined [1]{%
 \@ifx{#1\undefined}
}%
\providecommand \@ifnum [1]{%
 \ifnum #1\expandafter \@firstoftwo
 \else \expandafter \@secondoftwo
 \fi
}%
\providecommand \@ifx [1]{%
 \ifx #1\expandafter \@firstoftwo
 \else \expandafter \@secondoftwo
 \fi
}%
\providecommand \natexlab [1]{#1}%
\providecommand \enquote  [1]{``#1''}%
\providecommand \bibnamefont  [1]{#1}%
\providecommand \bibfnamefont [1]{#1}%
\providecommand \citenamefont [1]{#1}%
\providecommand \href@noop [0]{\@secondoftwo}%
\providecommand \href [0]{\begingroup \@sanitize@url \@href}%
\providecommand \@href[1]{\@@startlink{#1}\@@href}%
\providecommand \@@href[1]{\endgroup#1\@@endlink}%
\providecommand \@sanitize@url [0]{\catcode `\\12\catcode `\$12\catcode
  `\&12\catcode `\#12\catcode `\^12\catcode `\_12\catcode `\%12\relax}%
\providecommand \@@startlink[1]{}%
\providecommand \@@endlink[0]{}%
\providecommand \url  [0]{\begingroup\@sanitize@url \@url }%
\providecommand \@url [1]{\endgroup\@href {#1}{\urlprefix }}%
\providecommand \urlprefix  [0]{URL }%
\providecommand \Eprint [0]{\href }%
\providecommand \doibase [0]{https://doi.org/}%
\providecommand \selectlanguage [0]{\@gobble}%
\providecommand \bibinfo  [0]{\@secondoftwo}%
\providecommand \bibfield  [0]{\@secondoftwo}%
\providecommand \translation [1]{[#1]}%
\providecommand \BibitemOpen [0]{}%
\providecommand \bibitemStop [0]{}%
\providecommand \bibitemNoStop [0]{.\EOS\space}%
\providecommand \EOS [0]{\spacefactor3000\relax}%
\providecommand \BibitemShut  [1]{\csname bibitem#1\endcsname}%
\let\auto@bib@innerbib\@empty
\bibitem [{\citenamefont {Kikuchi}\ \emph
  {et~al.}(2005{\natexlab{a}})\citenamefont {Kikuchi}, \citenamefont {Fujii},
  \citenamefont {Chiba}, \citenamefont {Mitsudo}, \citenamefont {Idehara},
  \citenamefont {Tonegawa}, \citenamefont {Okamoto}, \citenamefont {Sakai},
  \citenamefont {Kuwai},\ and\ \citenamefont {Ohta}}]{Kikuchi2005-prl}%
  \BibitemOpen
  \bibfield  {author} {\bibinfo {author} {\bibfnamefont {H.}~\bibnamefont
  {Kikuchi}}, \bibinfo {author} {\bibfnamefont {Y.}~\bibnamefont {Fujii}},
  \bibinfo {author} {\bibfnamefont {M.}~\bibnamefont {Chiba}}, \bibinfo
  {author} {\bibfnamefont {S.}~\bibnamefont {Mitsudo}}, \bibinfo {author}
  {\bibfnamefont {T.}~\bibnamefont {Idehara}}, \bibinfo {author} {\bibfnamefont
  {T.}~\bibnamefont {Tonegawa}}, \bibinfo {author} {\bibfnamefont
  {K.}~\bibnamefont {Okamoto}}, \bibinfo {author} {\bibfnamefont
  {T.}~\bibnamefont {Sakai}}, \bibinfo {author} {\bibfnamefont
  {T.}~\bibnamefont {Kuwai}},\ and\ \bibinfo {author} {\bibfnamefont
  {H.}~\bibnamefont {Ohta}},\ }\bibfield  {title} {\bibinfo {title}
  {Experimental {Observation} of the 1/3 {Magnetization} {Plateau} in the
  {Diamond}-{Chain} {Compound}
  {Cu}$_{\textrm{3}}$({CO}$_{\textrm{3}}$)$_{\textrm{2}}$({OH})$_{\textrm{2}}$},\
  }\href {https://doi.org/10.1103/PhysRevLett.94.227201} {\bibfield  {journal}
  {\bibinfo  {journal} {Physical Review Letters}\ }\textbf {\bibinfo {volume}
  {94}},\ \bibinfo {pages} {227201} (\bibinfo {year}
  {2005}{\natexlab{a}})}\BibitemShut {NoStop}%
\bibitem [{\citenamefont {Kikuchi}\ \emph
  {et~al.}(2005{\natexlab{b}})\citenamefont {Kikuchi}, \citenamefont {Fujii},
  \citenamefont {Chiba}, \citenamefont {Mitsudo}, \citenamefont {Idehara},
  \citenamefont {Tonegawa}, \citenamefont {Okamoto}, \citenamefont {Sakai},
  \citenamefont {Kuwai}, \citenamefont {Kindo}, \citenamefont {Matsuo},
  \citenamefont {Higemoto}, \citenamefont {Nishiyama}, \citenamefont
  {Horvatić},\ and\ \citenamefont {Bertheir}}]{Kikuchi2005}%
  \BibitemOpen
  \bibfield  {author} {\bibinfo {author} {\bibfnamefont {H.}~\bibnamefont
  {Kikuchi}}, \bibinfo {author} {\bibfnamefont {Y.}~\bibnamefont {Fujii}},
  \bibinfo {author} {\bibfnamefont {M.}~\bibnamefont {Chiba}}, \bibinfo
  {author} {\bibfnamefont {S.}~\bibnamefont {Mitsudo}}, \bibinfo {author}
  {\bibfnamefont {T.}~\bibnamefont {Idehara}}, \bibinfo {author} {\bibfnamefont
  {T.}~\bibnamefont {Tonegawa}}, \bibinfo {author} {\bibfnamefont
  {K.}~\bibnamefont {Okamoto}}, \bibinfo {author} {\bibfnamefont
  {T.}~\bibnamefont {Sakai}}, \bibinfo {author} {\bibfnamefont
  {T.}~\bibnamefont {Kuwai}}, \bibinfo {author} {\bibfnamefont
  {K.}~\bibnamefont {Kindo}}, \bibinfo {author} {\bibfnamefont
  {A.}~\bibnamefont {Matsuo}}, \bibinfo {author} {\bibfnamefont
  {W.}~\bibnamefont {Higemoto}}, \bibinfo {author} {\bibfnamefont
  {K.}~\bibnamefont {Nishiyama}}, \bibinfo {author} {\bibfnamefont
  {M.}~\bibnamefont {Horvatić}},\ and\ \bibinfo {author} {\bibfnamefont
  {C.}~\bibnamefont {Bertheir}},\ }\bibfield  {title} {\bibinfo {title}
  {Magnetic {Properties} of the {Diamond} {Chain} {Compound}
  {Cu}$_{\textrm{3}}$({CO}$_{\textrm{3}}$)$_{\textrm{2}}$({OH})$_{\textrm{2}}$},\
  }\href {https://doi.org/10.1143/PTPS.159.1} {\bibfield  {journal} {\bibinfo
  {journal} {Progress of Theoretical Physics Supplement}\ }\textbf {\bibinfo
  {volume} {159}},\ \bibinfo {pages} {1} (\bibinfo {year}
  {2005}{\natexlab{b}})}\BibitemShut {NoStop}%
\bibitem [{\citenamefont {Honecker}\ \emph {et~al.}(2011)\citenamefont
  {Honecker}, \citenamefont {Hu}, \citenamefont {Peters},\ and\ \citenamefont
  {Richter}}]{Honecker2011}%
  \BibitemOpen
  \bibfield  {author} {\bibinfo {author} {\bibfnamefont {A.}~\bibnamefont
  {Honecker}}, \bibinfo {author} {\bibfnamefont {S.}~\bibnamefont {Hu}},
  \bibinfo {author} {\bibfnamefont {R.}~\bibnamefont {Peters}},\ and\ \bibinfo
  {author} {\bibfnamefont {J.}~\bibnamefont {Richter}},\ }\bibfield  {title}
  {\bibinfo {title} {Dynamic and thermodynamic properties of the generalized
  diamond chain model for azurite},\ }\href
  {https://doi.org/10.1088/0953-8984/23/16/164211} {\bibfield  {journal}
  {\bibinfo  {journal} {Journal of Physics: Condensed Matter}\ }\textbf
  {\bibinfo {volume} {23}},\ \bibinfo {pages} {164211} (\bibinfo {year}
  {2011})}\BibitemShut {NoStop}%
\bibitem [{\citenamefont {Rojas}\ \emph
  {et~al.}(2012{\natexlab{a}})\citenamefont {Rojas}, \citenamefont {Rojas},
  \citenamefont {Ananikian},\ and\ \citenamefont {de~Souza}}]{Rojas2012-pra}%
  \BibitemOpen
  \bibfield  {author} {\bibinfo {author} {\bibfnamefont {O.}~\bibnamefont
  {Rojas}}, \bibinfo {author} {\bibfnamefont {M.}~\bibnamefont {Rojas}},
  \bibinfo {author} {\bibfnamefont {N.~S.}\ \bibnamefont {Ananikian}},\ and\
  \bibinfo {author} {\bibfnamefont {S.~M.}\ \bibnamefont {de~Souza}},\
  }\bibfield  {title} {\bibinfo {title} {Thermal entanglement in an exactly
  solvable {Ising}-\textit{{XXZ}} diamond chain structure},\ }\href
  {https://doi.org/10.1103/PhysRevA.86.042330} {\bibfield  {journal} {\bibinfo
  {journal} {Physical Review A}\ }\textbf {\bibinfo {volume} {86}},\ \bibinfo
  {pages} {042330} (\bibinfo {year} {2012}{\natexlab{a}})}\BibitemShut
  {NoStop}%
\bibitem [{\citenamefont {Rojas}\ \emph
  {et~al.}(2012{\natexlab{b}})\citenamefont {Rojas}, \citenamefont {de~Souza},\
  and\ \citenamefont {Ananikian}}]{Rojas2012}%
  \BibitemOpen
  \bibfield  {author} {\bibinfo {author} {\bibfnamefont {O.}~\bibnamefont
  {Rojas}}, \bibinfo {author} {\bibfnamefont {S.~M.}\ \bibnamefont
  {de~Souza}},\ and\ \bibinfo {author} {\bibfnamefont {N.~S.}\ \bibnamefont
  {Ananikian}},\ }\bibfield  {title} {\bibinfo {title} {Geometrical frustration
  of an extended {Hubbard} diamond chain in the quasiatomic limit},\ }\href
  {https://doi.org/10.1103/PhysRevE.85.061123} {\bibfield  {journal} {\bibinfo
  {journal} {Physical Review E}\ }\textbf {\bibinfo {volume} {85}},\ \bibinfo
  {pages} {061123} (\bibinfo {year} {2012}{\natexlab{b}})}\BibitemShut
  {NoStop}%
\bibitem [{\citenamefont {Ananikian}\ \emph {et~al.}(2012)\citenamefont
  {Ananikian}, \citenamefont {Ananikyan}, \citenamefont {Chakhmakhchyan},\ and\
  \citenamefont {Rojas}}]{Ananikian2012}%
  \BibitemOpen
  \bibfield  {author} {\bibinfo {author} {\bibfnamefont {N.~S.}\ \bibnamefont
  {Ananikian}}, \bibinfo {author} {\bibfnamefont {L.~N.}\ \bibnamefont
  {Ananikyan}}, \bibinfo {author} {\bibfnamefont {L.~A.}\ \bibnamefont
  {Chakhmakhchyan}},\ and\ \bibinfo {author} {\bibfnamefont {O.}~\bibnamefont
  {Rojas}},\ }\bibfield  {title} {\bibinfo {title} {Thermal entanglement of a
  spin-1/2 {Ising}–{Heisenberg} model on a symmetrical diamond chain},\
  }\href {https://doi.org/10.1088/0953-8984/24/25/256001} {\bibfield  {journal}
  {\bibinfo  {journal} {Journal of Physics: Condensed Matter}\ }\textbf
  {\bibinfo {volume} {24}},\ \bibinfo {pages} {256001} (\bibinfo {year}
  {2012})}\BibitemShut {NoStop}%
\bibitem [{\citenamefont {G\'{a}lisov\'{a}}(2013)}]{Galisova2013}%
  \BibitemOpen
  \bibfield  {author} {\bibinfo {author} {\bibfnamefont {L.}~\bibnamefont
  {G\'{a}lisov\'{a}}},\ }\bibfield  {title} {\bibinfo {title} {Magnetic properties of
  the spin-1/2 {Ising}-{Heisenberg} diamond chain with the four-spin
  interaction: {Diamond} chain with the four-spin interaction},\ }\href
  {https://doi.org/10.1002/pssb.201248260} {\bibfield  {journal} {\bibinfo
  {journal} {physica status solidi (b)}\ }\textbf {\bibinfo {volume} {250}},\
  \bibinfo {pages} {187} (\bibinfo {year} {2013})}\BibitemShut {NoStop}%
\bibitem [{\citenamefont {Torrico}\ \emph {et~al.}(2014)\citenamefont
  {Torrico}, \citenamefont {Rojas}, \citenamefont {de~Souza}, \citenamefont
  {Rojas},\ and\ \citenamefont {Ananikian}}]{Torrico2014}%
  \BibitemOpen
  \bibfield  {author} {\bibinfo {author} {\bibfnamefont {J.}~\bibnamefont
  {Torrico}}, \bibinfo {author} {\bibfnamefont {M.}~\bibnamefont {Rojas}},
  \bibinfo {author} {\bibfnamefont {S.~M.}\ \bibnamefont {de~Souza}}, \bibinfo
  {author} {\bibfnamefont {O.}~\bibnamefont {Rojas}},\ and\ \bibinfo {author}
  {\bibfnamefont {N.~S.}\ \bibnamefont {Ananikian}},\ }\bibfield  {title}
  {\bibinfo {title} {Pairwise thermal entanglement in the {Ising}-{XYZ} diamond
  chain structure in an external magnetic field},\ }\href
  {https://doi.org/10.1209/0295-5075/108/50007} {\bibfield  {journal} {\bibinfo
   {journal} {EPL (Europhysics Letters)}\ }\textbf {\bibinfo {volume} {108}},\
  \bibinfo {pages} {50007} (\bibinfo {year} {2014})}\BibitemShut {NoStop}%
\bibitem [{\citenamefont {Derzhko}\ \emph {et~al.}(2015)\citenamefont
  {Derzhko}, \citenamefont {Krupnitska}, \citenamefont {Lisnyi},\ and\
  \citenamefont {Stre{\v{c}}ka}}]{Derzhko2015}%
  \BibitemOpen
  \bibfield  {author} {\bibinfo {author} {\bibfnamefont {O.}~\bibnamefont
  {Derzhko}}, \bibinfo {author} {\bibfnamefont {O.}~\bibnamefont {Krupnitska}},
  \bibinfo {author} {\bibfnamefont {B.}~\bibnamefont {Lisnyi}},\ and\ \bibinfo
  {author} {\bibfnamefont {J.}~\bibnamefont {Stre{\v{c}}ka}},\ }\bibfield  {title}
  {\bibinfo {title} {Effective low-energy description of almost
  {Ising}-{Heisenberg} diamond chain},\ }\href
  {https://doi.org/10.1209/0295-5075/112/37002} {\bibfield  {journal} {\bibinfo
   {journal} {EPL (Europhysics Letters)}\ }\textbf {\bibinfo {volume} {112}},\
  \bibinfo {pages} {37002} (\bibinfo {year} {2015})}\BibitemShut {NoStop}%
\bibitem [{\citenamefont {Richter}\ \emph {et~al.}(2015)\citenamefont
  {Richter}, \citenamefont {Krupnitska}, \citenamefont {Krokhmalskii},\ and\
  \citenamefont {Derzhko}}]{Richter2015}%
  \BibitemOpen
  \bibfield  {author} {\bibinfo {author} {\bibfnamefont {J.}~\bibnamefont
  {Richter}}, \bibinfo {author} {\bibfnamefont {O.}~\bibnamefont {Krupnitska}},
  \bibinfo {author} {\bibfnamefont {T.}~\bibnamefont {Krokhmalskii}},\ and\
  \bibinfo {author} {\bibfnamefont {O.}~\bibnamefont {Derzhko}},\ }\bibfield
  {title} {\bibinfo {title} {Frustrated diamond-chain quantum {XXZ}
  {Heisenberg} antiferromagnet in a magnetic field},\ }\href
  {https://doi.org/10.1016/j.jmmm.2014.11.082} {\bibfield  {journal} {\bibinfo
  {journal} {Journal of Magnetism and Magnetic Materials}\ }\textbf {\bibinfo
  {volume} {379}},\ \bibinfo {pages} {39} (\bibinfo {year} {2015})}\BibitemShut
  {NoStop}%
\bibitem [{\citenamefont {Lisnyi}\ and\ \citenamefont
  {Stre{\v{c}}ka}(2015)}]{Lisnyi2015}%
  \BibitemOpen
  \bibfield  {author} {\bibinfo {author} {\bibfnamefont {B.}~\bibnamefont
  {Lisnyi}}\ and\ \bibinfo {author} {\bibfnamefont {J.}~\bibnamefont
  {Stre{\v{c}}ka}},\ }\bibfield  {title} {\bibinfo {title} {Exactly solved mixed
  spin-(1,1/2) {Ising}–{Heisenberg} diamond chain with a single-ion
  anisotropy},\ }\href {https://doi.org/10.1016/j.jmmm.2014.10.113} {\bibfield
  {journal} {\bibinfo  {journal} {Journal of Magnetism and Magnetic Materials}\
  }\textbf {\bibinfo {volume} {377}},\ \bibinfo {pages} {502} (\bibinfo {year}
  {2015})}\BibitemShut {NoStop}%
\bibitem [{\citenamefont {Torrico}\ \emph
  {et~al.}(2016{\natexlab{a}})\citenamefont {Torrico}, \citenamefont {Rojas},
  \citenamefont {de~Souza},\ and\ \citenamefont {Rojas}}]{Torrico2016}%
  \BibitemOpen
  \bibfield  {author} {\bibinfo {author} {\bibfnamefont {J.}~\bibnamefont
  {Torrico}}, \bibinfo {author} {\bibfnamefont {M.}~\bibnamefont {Rojas}},
  \bibinfo {author} {\bibfnamefont {S.}~\bibnamefont {de~Souza}},\ and\
  \bibinfo {author} {\bibfnamefont {O.}~\bibnamefont {Rojas}},\ }\bibfield
  {title} {\bibinfo {title} {Zero temperature non-plateau magnetization and
  magnetocaloric effect in an {Ising}-{XYZ} diamond chain structure},\ }\href
  {https://doi.org/10.1016/j.physleta.2016.08.007} {\bibfield  {journal}
  {\bibinfo  {journal} {Physics Letters A}\ }\textbf {\bibinfo {volume}
  {380}},\ \bibinfo {pages} {3655} (\bibinfo {year}
  {2016}{\natexlab{a}})}\BibitemShut {NoStop}%
\bibitem [{\citenamefont {Lisnyi}\ and\ \citenamefont
  {Stre{\v{c}}ka}(2016)}]{Lisnyi2016}%
  \BibitemOpen
  \bibfield  {author} {\bibinfo {author} {\bibfnamefont {B.}~\bibnamefont
  {Lisnyi}}\ and\ \bibinfo {author} {\bibfnamefont {J.}~\bibnamefont
  {Stre{\v{c}}ka}},\ }\bibfield  {title} {\bibinfo {title} {Exactly solved mixed
  spin-(1,1/2) {Ising}–{Heisenberg} distorted diamond chain},\ }\href
  {https://doi.org/10.1016/j.physa.2016.06.088} {\bibfield  {journal} {\bibinfo
   {journal} {Physica A: Statistical Mechanics and its Applications}\ }\textbf
  {\bibinfo {volume} {462}},\ \bibinfo {pages} {104} (\bibinfo {year}
  {2016})}\BibitemShut {NoStop}%
\bibitem [{\citenamefont {Hovhannisyan}\ \emph {et~al.}(2016)\citenamefont
  {Hovhannisyan}, \citenamefont {Stre{\v{c}}ka},\ and\ \citenamefont
  {Ananikian}}]{Hovhannisyan2016}%
  \BibitemOpen
  \bibfield  {author} {\bibinfo {author} {\bibfnamefont {V.~V.}\ \bibnamefont
  {Hovhannisyan}}, \bibinfo {author} {\bibfnamefont {J.}~\bibnamefont
  {Stre{\v{c}}ka}},\ and\ \bibinfo {author} {\bibfnamefont {N.~S.}\ \bibnamefont
  {Ananikian}},\ }\bibfield  {title} {\bibinfo {title} {Exactly solvable spin-1
  {Ising}–{Heisenberg} diamond chain with the second-neighbor interaction
  between nodal spins},\ }\href {https://doi.org/10.1088/0953-8984/28/8/085401}
  {\bibfield  {journal} {\bibinfo  {journal} {Journal of Physics: Condensed
  Matter}\ }\textbf {\bibinfo {volume} {28}},\ \bibinfo {pages} {085401}
  (\bibinfo {year} {2016})}\BibitemShut {NoStop}%
\bibitem [{\citenamefont {Torrico}\ \emph
  {et~al.}(2016{\natexlab{b}})\citenamefont {Torrico}, \citenamefont {Rojas},
  \citenamefont {Pereira}, \citenamefont {Stre{\v{c}}ka},\ and\ \citenamefont
  {Lyra}}]{Torrico2016-prb}%
  \BibitemOpen
  \bibfield  {author} {\bibinfo {author} {\bibfnamefont {J.}~\bibnamefont
  {Torrico}}, \bibinfo {author} {\bibfnamefont {M.}~\bibnamefont {Rojas}},
  \bibinfo {author} {\bibfnamefont {M.~S.~S.}\ \bibnamefont {Pereira}},
  \bibinfo {author} {\bibfnamefont {J.}~\bibnamefont {Stre{\v{c}}ka}},\ and\
  \bibinfo {author} {\bibfnamefont {M.~L.}\ \bibnamefont {Lyra}},\ }\bibfield
  {title} {\bibinfo {title} {Spin frustration and fermionic entanglement in an
  exactly solved hybrid diamond chain with localized {Ising} spins and mobile
  electrons},\ }\href {https://doi.org/10.1103/PhysRevB.93.014428} {\bibfield
  {journal} {\bibinfo  {journal} {Physical Review B}\ }\textbf {\bibinfo
  {volume} {93}},\ \bibinfo {pages} {014428} (\bibinfo {year}
  {2016}{\natexlab{b}})}\BibitemShut {NoStop}%
\bibitem [{\citenamefont {Carvalho}\ \emph {et~al.}(2018)\citenamefont
  {Carvalho}, \citenamefont {Torrico}, \citenamefont {de~Souza}, \citenamefont
  {Rojas},\ and\ \citenamefont {Rojas}}]{Carvalho2018}%
  \BibitemOpen
  \bibfield  {author} {\bibinfo {author} {\bibfnamefont {I.}~\bibnamefont
  {Carvalho}}, \bibinfo {author} {\bibfnamefont {J.}~\bibnamefont {Torrico}},
  \bibinfo {author} {\bibfnamefont {S.}~\bibnamefont {de~Souza}}, \bibinfo
  {author} {\bibfnamefont {M.}~\bibnamefont {Rojas}},\ and\ \bibinfo {author}
  {\bibfnamefont {O.}~\bibnamefont {Rojas}},\ }\bibfield  {title} {\bibinfo
  {title} {Quantum entanglement in the neighborhood of pseudo-transition for a
  spin-1/2 {Ising}-{XYZ} diamond chain},\ }\href
  {https://doi.org/10.1016/j.jmmm.2018.06.018} {\bibfield  {journal} {\bibinfo
  {journal} {Journal of Magnetism and Magnetic Materials}\ }\textbf {\bibinfo
  {volume} {465}},\ \bibinfo {pages} {323} (\bibinfo {year}
  {2018})}\BibitemShut {NoStop}%
\bibitem [{\citenamefont {Carvalho}\ \emph {et~al.}(2019)\citenamefont
  {Carvalho}, \citenamefont {Torrico}, \citenamefont {de~Souza}, \citenamefont
  {Rojas},\ and\ \citenamefont {Derzhko}}]{Carvalho2019}%
  \BibitemOpen
  \bibfield  {author} {\bibinfo {author} {\bibfnamefont {I.}~\bibnamefont
  {Carvalho}}, \bibinfo {author} {\bibfnamefont {J.}~\bibnamefont {Torrico}},
  \bibinfo {author} {\bibfnamefont {S.}~\bibnamefont {de~Souza}}, \bibinfo
  {author} {\bibfnamefont {O.}~\bibnamefont {Rojas}},\ and\ \bibinfo {author}
  {\bibfnamefont {O.}~\bibnamefont {Derzhko}},\ }\bibfield  {title} {\bibinfo
  {title} {Correlation functions for a spin-1/2 {Ising}-{XYZ} diamond chain:
  {Further} evidence for quasi-phases and pseudo-transitions},\ }\href
  {https://doi.org/10.1016/j.aop.2019.01.001} {\bibfield  {journal} {\bibinfo
  {journal} {Annals of Physics}\ }\textbf {\bibinfo {volume} {402}},\ \bibinfo
  {pages} {45} (\bibinfo {year} {2019})}\BibitemShut {NoStop}%
\bibitem [{\citenamefont {Krokhmalskii}\ \emph {et~al.}(2021)\citenamefont
  {Krokhmalskii}, \citenamefont {Hutak}, \citenamefont {Rojas}, \citenamefont
  {de~Souza},\ and\ \citenamefont {Derzhko}}]{Krokhmalskii2021}%
  \BibitemOpen
  \bibfield  {author} {\bibinfo {author} {\bibfnamefont {T.}~\bibnamefont
  {Krokhmalskii}}, \bibinfo {author} {\bibfnamefont {T.}~\bibnamefont {Hutak}},
  \bibinfo {author} {\bibfnamefont {O.}~\bibnamefont {Rojas}}, \bibinfo
  {author} {\bibfnamefont {S.~M.}\ \bibnamefont {de~Souza}},\ and\ \bibinfo
  {author} {\bibfnamefont {O.}~\bibnamefont {Derzhko}},\ }\bibfield  {title}
  {\bibinfo {title} {Towards low-temperature peculiarities of thermodynamic
  quantities for decorated spin chains},\ }\href
  {https://doi.org/10.1016/j.physa.2021.125986} {\bibfield  {journal} {\bibinfo
   {journal} {Physica A: Statistical Mechanics and its Applications}\ }\textbf
  {\bibinfo {volume} {573}},\ \bibinfo {pages} {125986} (\bibinfo {year}
  {2021})}\BibitemShut {NoStop}%
\bibitem [{\citenamefont {Rojas}\ \emph {et~al.}(2021)\citenamefont {Rojas},
  \citenamefont {de~Souza}, \citenamefont {Torrico}, \citenamefont
  {Verissimo}, \citenamefont {Pereira},\ and\ \citenamefont
  {Lyra}}]{Rojas2021}%
  \BibitemOpen
  \bibfield  {author} {\bibinfo {author} {\bibfnamefont {O.}~\bibnamefont
  {Rojas}}, \bibinfo {author} {\bibfnamefont {S.~M.}\ \bibnamefont {de~Souza}},
  \bibinfo {author} {\bibfnamefont {J.}~\bibnamefont {Torrico}}, \bibinfo
  {author} {\bibfnamefont {L.~M.}\ \bibnamefont {Verissimo}}, \bibinfo
  {author} {\bibfnamefont {M.~S.~S.}\ \bibnamefont {Pereira}},\ and\ \bibinfo
  {author} {\bibfnamefont {M.~L.}\ \bibnamefont {Lyra}},\ }\bibfield  {title}
  {\bibinfo {title} {Low-temperature pseudo-phase-transition in an extended
  {Hubbard} diamond chain},\ }\href
  {https://doi.org/10.1103/PhysRevE.103.042123} {\bibfield  {journal} {\bibinfo
   {journal} {Physical Review E}\ }\textbf {\bibinfo {volume} {103}},\ \bibinfo
  {pages} {042123} (\bibinfo {year} {2021})}\BibitemShut {NoStop}%
\bibitem [{\citenamefont {Antonosyan}\ \emph {et~al.}(2009)\citenamefont
  {Antonosyan}, \citenamefont {Bellucci},\ and\ \citenamefont
  {Ohanyan}}]{Antonosyan2009}%
  \BibitemOpen
  \bibfield  {author} {\bibinfo {author} {\bibfnamefont {D.}~\bibnamefont
  {Antonosyan}}, \bibinfo {author} {\bibfnamefont {S.}~\bibnamefont
  {Bellucci}},\ and\ \bibinfo {author} {\bibfnamefont {V.}~\bibnamefont
  {Ohanyan}},\ }\bibfield  {title} {\bibinfo {title} {Exactly solvable
  {Ising}-{Heisenberg} chain with triangular {X}{X}{Z} -{Heisenberg}
  plaquettes},\ }\href {https://doi.org/10.1103/PhysRevB.79.014432} {\bibfield
  {journal} {\bibinfo  {journal} {Physical Review B}\ }\textbf {\bibinfo
  {volume} {79}},\ \bibinfo {pages} {014432} (\bibinfo {year}
  {2009})}\BibitemShut {NoStop}%
\bibitem [{\citenamefont {G\'{a}lisov\'{a}}\ and\ \citenamefont
  {Stre{\v{c}}ka}(2015{\natexlab{a}})}]{Galisova2015-pre}%
  \BibitemOpen
  \bibfield  {author} {\bibinfo {author} {\bibfnamefont {L.}~\bibnamefont
  {G\'{a}lisov\'{a}}}\ and\ \bibinfo {author} {\bibfnamefont {J.}~\bibnamefont
  {Stre{\v{c}}ka}},\ }\bibfield  {title} {\bibinfo {title} {Vigorous thermal
  excitations in a double-tetrahedral chain of localized {Ising} spins and
  mobile electrons mimic a temperature-driven first-order phase transition},\
  }\href {https://doi.org/10.1103/PhysRevE.91.022134} {\bibfield  {journal}
  {\bibinfo  {journal} {Physical Review E}\ }\textbf {\bibinfo {volume} {91}},\
  \bibinfo {pages} {022134} (\bibinfo {year} {2015}{\natexlab{a}})}\BibitemShut
  {NoStop}%
\bibitem [{\citenamefont {G\'{a}lisov\'{a}}\ and\ \citenamefont
  {Stre{\v{c}}ka}(2015{\natexlab{b}})}]{Galisova2015}%
  \BibitemOpen
  \bibfield  {author} {\bibinfo {author} {\bibfnamefont {L.}~\bibnamefont
  {G\'{a}lisov\'{a}}}\ and\ \bibinfo {author} {\bibfnamefont {J.}~\bibnamefont
  {Stre{\v{c}}ka}},\ }\bibfield  {title} {\bibinfo {title} {Magnetic {Grüneisen}
  parameter and magnetocaloric properties of a coupled spin–electron
  double-tetrahedral chain},\ }\href
  {https://doi.org/10.1016/j.physleta.2015.07.007} {\bibfield  {journal}
  {\bibinfo  {journal} {Physics Letters A}\ }\textbf {\bibinfo {volume}
  {379}},\ \bibinfo {pages} {2474} (\bibinfo {year}
  {2015}{\natexlab{b}})}\BibitemShut {NoStop}%
\bibitem [{\citenamefont {de~Souza}\ and\ \citenamefont
  {Rojas}(2018)}]{DeSouza2018}%
  \BibitemOpen
  \bibfield  {author} {\bibinfo {author} {\bibfnamefont {S.}~\bibnamefont
  {de~Souza}}\ and\ \bibinfo {author} {\bibfnamefont {O.}~\bibnamefont
  {Rojas}},\ }\bibfield  {title} {\bibinfo {title} {Quasi-phases and
  pseudo-transitions in one-dimensional models with nearest neighbor
  interactions},\ }\href {https://doi.org/10.1016/j.ssc.2017.10.006} {\bibfield
   {journal} {\bibinfo  {journal} {Solid State Communications}\ }\textbf
  {\bibinfo {volume} {269}},\ \bibinfo {pages} {131} (\bibinfo {year}
  {2018})}\BibitemShut {NoStop}%
\bibitem [{\citenamefont {G\'{a}lisov\'{a}}\ and\ \citenamefont
  {Knežo}(2018)}]{Galisova2018}%
  \BibitemOpen
  \bibfield  {author} {\bibinfo {author} {\bibfnamefont {L.}~\bibnamefont
  {G\'{a}lisov\'{a}}}\ and\ \bibinfo {author} {\bibfnamefont {D.}~\bibnamefont
  {Knežo}},\ }\bibfield  {title} {\bibinfo {title} {Macroscopic ground-state
  degeneracy and magnetocaloric effect in the exactly solvable spin-1/2
  {Ising}–{Heisenberg} double-tetrahedral chain},\ }\href
  {https://doi.org/10.1016/j.physleta.2018.06.012} {\bibfield  {journal}
  {\bibinfo  {journal} {Physics Letters A}\ }\textbf {\bibinfo {volume}
  {382}},\ \bibinfo {pages} {2839} (\bibinfo {year} {2018})}\BibitemShut
  {NoStop}%
\bibitem [{\citenamefont {Stre{\v{c}}ka}\ \emph {et~al.}(2014)\citenamefont
  {Stre{\v{c}}ka}, \citenamefont {Rojas}, \citenamefont {Verkholyak},\ and\
  \citenamefont {Lyra}}]{Strecka2014}%
  \BibitemOpen
  \bibfield  {author} {\bibinfo {author} {\bibfnamefont {J.}~\bibnamefont
  {Stre{\v{c}}ka}}, \bibinfo {author} {\bibfnamefont {O.}~\bibnamefont {Rojas}},
  \bibinfo {author} {\bibfnamefont {T.}~\bibnamefont {Verkholyak}},\ and\
  \bibinfo {author} {\bibfnamefont {M.~L.}\ \bibnamefont {Lyra}},\ }\bibfield
  {title} {\bibinfo {title} {Magnetization process, bipartite entanglement, and
  enhanced magnetocaloric effect of the exactly solved spin-1/2
  {Ising}-{Heisenberg} tetrahedral chain},\ }\href
  {https://doi.org/10.1103/PhysRevE.89.022143} {\bibfield  {journal} {\bibinfo
  {journal} {Physical Review E}\ }\textbf {\bibinfo {volume} {89}},\ \bibinfo
  {pages} {022143} (\bibinfo {year} {2014})}\BibitemShut {NoStop}%
\bibitem [{\citenamefont {Rojas}\ \emph {et~al.}(2016)\citenamefont {Rojas},
  \citenamefont {Stre{\v{c}}ka},\ and\ \citenamefont {de~Souza}}]{Rojas2016}%
  \BibitemOpen
  \bibfield  {author} {\bibinfo {author} {\bibfnamefont {O.}~\bibnamefont
  {Rojas}}, \bibinfo {author} {\bibfnamefont {J.}~\bibnamefont {Stre{\v{c}}ka}},\
  and\ \bibinfo {author} {\bibfnamefont {S.}~\bibnamefont {de~Souza}},\
  }\bibfield  {title} {\bibinfo {title} {Thermal entanglement and sharp
  specific-heat peak in an exactly solved spin-1/2 {Ising}-{Heisenberg} ladder
  with alternating {Ising} and {Heisenberg} inter–leg couplings},\ }\href
  {https://doi.org/10.1016/j.ssc.2016.08.002} {\bibfield  {journal} {\bibinfo
  {journal} {Solid State Communications}\ }\textbf {\bibinfo {volume} {246}},\
  \bibinfo {pages} {68} (\bibinfo {year} {2016})}\BibitemShut {NoStop}%
\bibitem [{\citenamefont {Sousa}\ \emph {et~al.}(2018)\citenamefont {Sousa},
  \citenamefont {Pereira}, \citenamefont {de~Oliveira}, \citenamefont
  {Stre{\v{c}}ka},\ and\ \citenamefont {Lyra}}]{Sousa2018}%
  \BibitemOpen
  \bibfield  {author} {\bibinfo {author} {\bibfnamefont {H.~S.}\ \bibnamefont
  {Sousa}}, \bibinfo {author} {\bibfnamefont {M.~S.~S.}\ \bibnamefont
  {Pereira}}, \bibinfo {author} {\bibfnamefont {I.~N.}\ \bibnamefont
  {de~Oliveira}}, \bibinfo {author} {\bibfnamefont {J.}~\bibnamefont
  {Stre{\v{c}}ka}},\ and\ \bibinfo {author} {\bibfnamefont {M.~L.}\ \bibnamefont
  {Lyra}},\ }\bibfield  {title} {\bibinfo {title} {Phase diagram and re-entrant
  fermionic entanglement in a hybrid {Ising}-{Hubbard} ladder},\ }\href
  {https://doi.org/10.1103/PhysRevE.97.052115} {\bibfield  {journal} {\bibinfo
  {journal} {Physical Review E}\ }\textbf {\bibinfo {volume} {97}},\ \bibinfo
  {pages} {052115} (\bibinfo {year} {2018})}\BibitemShut {NoStop}%
\bibitem [{\citenamefont {Al\'{e}cio}\ \emph {et~al.}(2016)\citenamefont
  {Al\'{e}cio}, \citenamefont {Lyra},\ and\ \citenamefont
  {Stre{\v{c}}ka}}]{Alecio2016}%
  \BibitemOpen
  \bibfield  {author} {\bibinfo {author} {\bibfnamefont {R.~C.}\ \bibnamefont
  {Al\'{e}cio}}, \bibinfo {author} {\bibfnamefont {M.~L.}\ \bibnamefont {Lyra}},\
  and\ \bibinfo {author} {\bibfnamefont {J.}~\bibnamefont {Stre{\v{c}}ka}},\
  }\bibfield  {title} {\bibinfo {title} {Ground states, magnetization plateaus
  and bipartite entanglement of frustrated spin-1/2 {Ising}-{Heisenberg} and
  {Heisenberg} triangular tubes},\ }\href
  {https://doi.org/10.1016/j.jmmm.2016.05.081} {\bibfield  {journal} {\bibinfo
  {journal} {Journal of Magnetism and Magnetic Materials}\ }\textbf {\bibinfo
  {volume} {417}},\ \bibinfo {pages} {294} (\bibinfo {year}
  {2016})}\BibitemShut {NoStop}%
\bibitem [{\citenamefont {Stre{\v{c}}ka}\ \emph {et~al.}(2016)\citenamefont
  {Stre{\v{c}}ka}, \citenamefont {Al\'{e}cio}, \citenamefont {Lyra},\ and\
  \citenamefont {Rojas}}]{Strecka2016}%
  \BibitemOpen
  \bibfield  {author} {\bibinfo {author} {\bibfnamefont {J.}~\bibnamefont
  {Stre{\v{c}}ka}}, \bibinfo {author} {\bibfnamefont {R.~C.}\ \bibnamefont
  {Al\'{e}cio}}, \bibinfo {author} {\bibfnamefont {M.~L.}\ \bibnamefont {Lyra}},\
  and\ \bibinfo {author} {\bibfnamefont {O.}~\bibnamefont {Rojas}},\ }\bibfield
   {title} {\bibinfo {title} {Spin frustration of a spin-1/2
  {Ising}–{Heisenberg} three-leg tube as an indispensable ground for thermal
  entanglement},\ }\href {https://doi.org/10.1016/j.jmmm.2016.02.095}
  {\bibfield  {journal} {\bibinfo  {journal} {Journal of Magnetism and Magnetic
  Materials}\ }\textbf {\bibinfo {volume} {409}},\ \bibinfo {pages} {124}
  (\bibinfo {year} {2016})}\BibitemShut {NoStop}%
\bibitem [{\citenamefont {Stre{\v{c}}ka}\ and\ \citenamefont
  {Dan{\v{c}}o}(2011)}]{Strecka2011}%
  \BibitemOpen
  \bibfield  {author} {\bibinfo {author} {\bibfnamefont {J.}~\bibnamefont
  {Stre{\v{c}}ka}}\ and\ \bibinfo {author} {\bibfnamefont {M.}~\bibnamefont
  {Dan{\v{c}}o}},\ }\bibfield  {title} {\bibinfo {title} {Unusual field-induced
  transitions in exactly solved mixed spin-(1/2, 1) {Ising} chain with axial
  and rhombic zero-field splitting parameters},\ }\href
  {https://doi.org/10.1016/j.physb.2011.04.040} {\bibfield  {journal} {\bibinfo
   {journal} {Physica B: Condensed Matter}\ }\textbf {\bibinfo {volume}
  {406}},\ \bibinfo {pages} {2967} (\bibinfo {year} {2011})}\BibitemShut
  {NoStop}%
\bibitem [{\citenamefont {Bellucci}\ \emph {et~al.}(2014)\citenamefont
  {Bellucci}, \citenamefont {Ohanyan},\ and\ \citenamefont
  {Rojas}}]{Bellucci2014}%
  \BibitemOpen
  \bibfield  {author} {\bibinfo {author} {\bibfnamefont {S.}~\bibnamefont
  {Bellucci}}, \bibinfo {author} {\bibfnamefont {V.}~\bibnamefont {Ohanyan}},\
  and\ \bibinfo {author} {\bibfnamefont {O.}~\bibnamefont {Rojas}},\ }\bibfield
   {title} {\bibinfo {title} {Magnetization non-rational quasi-plateau and
  spatially modulated spin order in the model of the single-chain magnet,
  [\{({CuL})$_{\textrm{2}}${Dy}\}\{{Mo}({CN})$_{\textrm{8}}$\}]{2CH}$_{\textrm{3}}${CN}{H}$_{\textrm{2}}${O}},\
  }\href {https://doi.org/10.1209/0295-5075/105/47012} {\bibfield  {journal}
  {\bibinfo  {journal} {EPL (Europhysics Letters)}\ }\textbf {\bibinfo {volume}
  {105}},\ \bibinfo {pages} {47012} (\bibinfo {year} {2014})}\BibitemShut
  {NoStop}%
\bibitem [{\citenamefont {Torrico}\ \emph {et~al.}(2018)\citenamefont
  {Torrico}, \citenamefont {Stre{\v{c}}ka}, \citenamefont {Hagiwara}, \citenamefont
  {Rojas}, \citenamefont {de~Souza}, \citenamefont {Han}, \citenamefont
  {Honda},\ and\ \citenamefont {Lyra}}]{Torrico2018}%
  \BibitemOpen
  \bibfield  {author} {\bibinfo {author} {\bibfnamefont {J.}~\bibnamefont
  {Torrico}}, \bibinfo {author} {\bibfnamefont {J.}~\bibnamefont {Stre{\v{c}}ka}},
  \bibinfo {author} {\bibfnamefont {M.}~\bibnamefont {Hagiwara}}, \bibinfo
  {author} {\bibfnamefont {O.}~\bibnamefont {Rojas}}, \bibinfo {author}
  {\bibfnamefont {S.}~\bibnamefont {de~Souza}}, \bibinfo {author}
  {\bibfnamefont {Y.}~\bibnamefont {Han}}, \bibinfo {author} {\bibfnamefont
  {Z.}~\bibnamefont {Honda}},\ and\ \bibinfo {author} {\bibfnamefont
  {M.}~\bibnamefont {Lyra}},\ }\bibfield  {title} {\bibinfo {title}
  {Heterobimetallic {Dy}-{Cu} coordination compound as a classical-quantum
  ferrimagnetic chain of regularly alternating {Ising} and {Heisenberg}
  spins},\ }\href {https://doi.org/10.1016/j.jmmm.2018.04.021} {\bibfield
  {journal} {\bibinfo  {journal} {Journal of Magnetism and Magnetic Materials}\
  }\textbf {\bibinfo {volume} {460}},\ \bibinfo {pages} {368} (\bibinfo {year}
  {2018})}\BibitemShut {NoStop}%
\bibitem [{\citenamefont {Verkholyak}\ and\ \citenamefont
  {Stre{\v{c}}ka}(2021)}]{Verkholyak2021}%
  \BibitemOpen
  \bibfield  {author} {\bibinfo {author} {\bibfnamefont {T.}~\bibnamefont
  {Verkholyak}}\ and\ \bibinfo {author} {\bibfnamefont {J.}~\bibnamefont
  {Stre{\v{c}}ka}},\ }\bibfield  {title} {\bibinfo {title} {Modified strong-coupling
  treatment of a spin-1/2 {Heisenberg} trimerized chain developed from the
  exactly solved {Ising}-{Heisenberg} diamond chain},\ }\href
  {https://doi.org/10.1103/PhysRevB.103.184415} {\bibfield  {journal} {\bibinfo
   {journal} {Physical Review B}\ }\textbf {\bibinfo {volume} {103}},\ \bibinfo
  {pages} {184415} (\bibinfo {year} {2021})}\BibitemShut {NoStop}%
\bibitem [{\citenamefont {Zhitomirsky}(2003)}]{Zhitomirsky2003}%
  \BibitemOpen
  \bibfield  {author} {\bibinfo {author} {\bibfnamefont {M.~E.}\ \bibnamefont
  {Zhitomirsky}},\ }\bibfield  {title} {\bibinfo {title} {Enhanced
  magnetocaloric effect in frustrated magnets},\ }\href
  {https://doi.org/10.1103/PhysRevB.67.104421} {\bibfield  {journal} {\bibinfo
  {journal} {Physical Review B}\ }\textbf {\bibinfo {volume} {67}},\ \bibinfo
  {pages} {104421} (\bibinfo {year} {2003})}\BibitemShut {NoStop}%
\bibitem [{\citenamefont {Zhitomirsky}\ and\ \citenamefont
  {Honecker}(2004)}]{Zhitomirsky2004}%
  \BibitemOpen
  \bibfield  {author} {\bibinfo {author} {\bibfnamefont {M.~E.}\ \bibnamefont
  {Zhitomirsky}}\ and\ \bibinfo {author} {\bibfnamefont {A.}~\bibnamefont
  {Honecker}},\ }\bibfield  {title} {\bibinfo {title} {Magnetocaloric effect in
  one-dimensional antiferromagnets},\ }\href
  {https://doi.org/10.1088/1742-5468/2004/07/P07012} {\bibfield  {journal}
  {\bibinfo  {journal} {Journal of Statistical Mechanics: Theory and
  Experiment}\ }\textbf {\bibinfo {volume} {2004}},\ \bibinfo {pages} {P07012}
  (\bibinfo {year} {2004})}\BibitemShut {NoStop}%
\bibitem [{\citenamefont {Sosin}\ \emph {et~al.}(2005)\citenamefont {Sosin},
  \citenamefont {Prozorova}, \citenamefont {Smirnov}, \citenamefont {Golov},
  \citenamefont {Berkutov}, \citenamefont {Petrenko}, \citenamefont
  {Balakrishnan},\ and\ \citenamefont {Zhitomirsky}}]{Sosin2005}%
  \BibitemOpen
  \bibfield  {author} {\bibinfo {author} {\bibfnamefont {S.~S.}\ \bibnamefont
  {Sosin}}, \bibinfo {author} {\bibfnamefont {L.~A.}\ \bibnamefont
  {Prozorova}}, \bibinfo {author} {\bibfnamefont {A.~I.}\ \bibnamefont
  {Smirnov}}, \bibinfo {author} {\bibfnamefont {A.~I.}\ \bibnamefont {Golov}},
  \bibinfo {author} {\bibfnamefont {I.~B.}\ \bibnamefont {Berkutov}}, \bibinfo
  {author} {\bibfnamefont {O.~A.}\ \bibnamefont {Petrenko}}, \bibinfo {author}
  {\bibfnamefont {G.}~\bibnamefont {Balakrishnan}},\ and\ \bibinfo {author}
  {\bibfnamefont {M.~E.}\ \bibnamefont {Zhitomirsky}},\ }\bibfield  {title}
  {\bibinfo {title} {Magnetocaloric effect in pyrochlore antiferromagnet
  {Gd}$_2${Ti}$_2${O}$_7$},\ }\href
  {https://doi.org/10.1103/PhysRevB.71.094413} {\bibfield  {journal} {\bibinfo
  {journal} {Physical Review B}\ }\textbf {\bibinfo {volume} {71}},\ \bibinfo
  {pages} {094413} (\bibinfo {year} {2005})}\BibitemShut {NoStop}%
\bibitem [{\citenamefont {Pereira}\ \emph {et~al.}(2009)\citenamefont
  {Pereira}, \citenamefont {de~Moura},\ and\ \citenamefont
  {Lyra}}]{Pereira2009}%
  \BibitemOpen
  \bibfield  {author} {\bibinfo {author} {\bibfnamefont {M.~S.~S.}\
  \bibnamefont {Pereira}}, \bibinfo {author} {\bibfnamefont {F.~A. B.~F.}\
  \bibnamefont {de~Moura}},\ and\ \bibinfo {author} {\bibfnamefont {M.~L.}\
  \bibnamefont {Lyra}},\ }\bibfield  {title} {\bibinfo {title} {Magnetocaloric
  effect in kinetically frustrated diamond chains},\ }\href
  {https://doi.org/10.1103/PhysRevB.79.054427} {\bibfield  {journal} {\bibinfo
  {journal} {Physical Review B}\ }\textbf {\bibinfo {volume} {79}},\ \bibinfo
  {pages} {054427} (\bibinfo {year} {2009})}\BibitemShut {NoStop}%
\bibitem [{\citenamefont {Rojas}\ \emph {et~al.}(2019)\citenamefont {Rojas},
  \citenamefont {Stre{\v{c}}ka}, \citenamefont {Lyra},\ and\ \citenamefont
  {de~Souza}}]{Rojas2019}%
  \BibitemOpen
  \bibfield  {author} {\bibinfo {author} {\bibfnamefont {O.}~\bibnamefont
  {Rojas}}, \bibinfo {author} {\bibfnamefont {J.}~\bibnamefont {Stre{\v{c}}ka}},
  \bibinfo {author} {\bibfnamefont {M.~L.}\ \bibnamefont {Lyra}},\ and\
  \bibinfo {author} {\bibfnamefont {S.~M.}\ \bibnamefont {de~Souza}},\
  }\bibfield  {title} {\bibinfo {title} {Universality and quasicritical
  exponents of one-dimensional models displaying a quasitransition at finite
  temperatures},\ }\href {https://doi.org/10.1103/PhysRevE.99.042117}
  {\bibfield  {journal} {\bibinfo  {journal} {Physical Review E}\ }\textbf
  {\bibinfo {volume} {99}},\ \bibinfo {pages} {042117} (\bibinfo {year}
  {2019})}\BibitemShut {NoStop}%
\bibitem [{\citenamefont {Rojas}(2020{\natexlab{a}})}]{Rojas2020-APP}%
  \BibitemOpen
  \bibfield  {author} {\bibinfo {author} {\bibfnamefont {O.}~\bibnamefont
  {Rojas}},\ }\bibfield  {title} {\bibinfo {title} {Residual {Entropy} and
  {Low} {Temperature} {Pseudo}-{Transition} for {One}-{Dimensional} {Models}},\
  }\href {https://doi.org/10.12693/APhysPolA.137.933} {\bibfield  {journal}
  {\bibinfo  {journal} {Acta Physica Polonica A}\ }\textbf {\bibinfo {volume}
  {137}},\ \bibinfo {pages} {933} (\bibinfo {year}
  {2020}{\natexlab{a}})}\BibitemShut {NoStop}%
\bibitem [{\citenamefont {Rojas}(2020{\natexlab{b}})}]{Rojas2020-BJP}%
  \BibitemOpen
  \bibfield  {author} {\bibinfo {author} {\bibfnamefont {O.}~\bibnamefont
  {Rojas}},\ }\bibfield  {title} {\bibinfo {title} {A {Conjecture} on the
  {Relationship} {Between} {Critical} {Residual} {Entropy} and {Finite}
  {Temperature} {Pseudo}-transitions of {One}-dimensional {Models}},\ }\href
  {https://doi.org/10.1007/s13538-020-00773-8} {\bibfield  {journal} {\bibinfo
  {journal} {Brazilian Journal of Physics}\ }\textbf {\bibinfo {volume} {50}},\
  \bibinfo {pages} {675} (\bibinfo {year} {2020}{\natexlab{b}})}\BibitemShut
  {NoStop}%
\bibitem [{\citenamefont {Katsura}\ and\ \citenamefont
  {Tsujiyama}(1965)}]{Katsura1965}%
  \BibitemOpen
  \bibfield  {author} {\bibinfo {author} {\bibfnamefont {S.}~\bibnamefont
  {Katsura}}\ and\ \bibinfo {author} {\bibfnamefont {B.}~\bibnamefont
  {Tsujiyama}},\ }\bibfield  {title} {\bibinfo {title} {{Ferro- and
  Antiferromagnetism of Dilute Ising Model}},\ }in\ \href@noop {} {\emph
  {\bibinfo {booktitle} {Proceedings of the Conference on PhenomenaPhenomena in
  the Neighborhood of Critical Points}}},\ \bibinfo {editor} {edited by\
  \bibinfo {editor} {\bibfnamefont {C.}~\bibnamefont {Domb}}}\ (\bibinfo
  {publisher} {National Bureau of Standards},\ \bibinfo {address} {Washington,
  D.C.},\ \bibinfo {year} {1965})\ pp.\ \bibinfo {pages} {219--224}\BibitemShut
  {NoStop}%
\bibitem [{\citenamefont {Rys}\ and\ \citenamefont
  {Hintermann}(1969)}]{Rys1969}%
  \BibitemOpen
  \bibfield  {author} {\bibfnamefont {A.}~\bibnamefont
  {Hintermann}}\ and\ \bibinfo {author} {\bibinfo {author} {\bibfnamefont {F.}~\bibnamefont
  {Rys}},\ }\bibfield  {title} {\bibinfo {title} {{Gittermodell eines ungeordneten Ferromagneten II. 
   Exakte L\"{o}sung des eindimensionalen Modells}},\
  }\href@noop {} {\bibfield  {journal} {\bibinfo  {journal} {Helv. Phys. Acta}\
  }\textbf {\bibinfo {volume} {42}},\ \bibinfo {pages} {608} (\bibinfo {year}
  {1969})}\BibitemShut {NoStop}%
\bibitem [{\citenamefont {Kawatra}\ and\ \citenamefont
  {Kijewski}(1969)}]{Kawatra1969}%
  \BibitemOpen
  \bibfield  {author} {\bibinfo {author} {\bibfnamefont {M.~P.}\ \bibnamefont
  {Kawatra}}\ and\ \bibinfo {author} {\bibfnamefont {L.~J.}\ \bibnamefont
  {Kijewski}},\ }\bibfield  {title} {\bibinfo {title} {Exact {Solution} of a
  {One}-{Dimensional} {Magnetic} {Lattice} {Gas} with an {Ising}
  {Interaction}},\ }\href {https://doi.org/10.1103/PhysRev.183.291} {\bibfield
  {journal} {\bibinfo  {journal} {Physical Review}\ }\textbf {\bibinfo {volume}
  {183}},\ \bibinfo {pages} {291} (\bibinfo {year} {1969})}\BibitemShut
  {NoStop}%
\bibitem [{\citenamefont {Matsubara}\ \emph {et~al.}(1973)\citenamefont
  {Matsubara}, \citenamefont {Yoshimura},\ and\ \citenamefont
  {Katsura}}]{Matsubara1973}%
  \BibitemOpen
  \bibfield  {author} {\bibinfo {author} {\bibfnamefont {F.}~\bibnamefont
  {Matsubara}}, \bibinfo {author} {\bibfnamefont {K.}~\bibnamefont
  {Yoshimura}},\ and\ \bibinfo {author} {\bibfnamefont {S.}~\bibnamefont
  {Katsura}},\ }\bibfield  {title} {\bibinfo {title} {Magnetic {Properties} of
  {One}-{Dimensional} {Dilute} {Ising} {Systems}. {I}},\ }\href
  {https://doi.org/10.1139/p73-140} {\bibfield  {journal} {\bibinfo  {journal}
  {Canadian Journal of Physics}\ }\textbf {\bibinfo {volume} {51}},\ \bibinfo
  {pages} {1053} (\bibinfo {year} {1973})}\BibitemShut {NoStop}%
\bibitem [{\citenamefont {Termonia}\ and\ \citenamefont
  {Deltour}(1974)}]{Termonia1974}%
  \BibitemOpen
  \bibfield  {author} {\bibinfo {author} {\bibfnamefont {Y.}~\bibnamefont
  {Termonia}}\ and\ \bibinfo {author} {\bibfnamefont {J.}~\bibnamefont
  {Deltour}},\ }\bibfield  {title} {\bibinfo {title} {Thermodynamic properties
  of one-dimensional dilute {Ising} systems with interacting impurities},\
  }\href {https://doi.org/10.1088/0022-3719/7/24/007} {\bibfield  {journal}
  {\bibinfo  {journal} {Journal of Physics C: Solid State Physics}\ }\textbf
  {\bibinfo {volume} {7}},\ \bibinfo {pages} {4441} (\bibinfo {year}
  {1974})}\BibitemShut {NoStop}%
\bibitem [{\citenamefont {Balagurov}\ \emph {et~al.}(1974)\citenamefont
  {Balagurov}, \citenamefont {Vaks},\ and\ \citenamefont
  {Zaitsev}}]{Balagurov1974}%
  \BibitemOpen
  \bibfield  {author} {\bibinfo {author} {\bibfnamefont {B.}~\bibnamefont
  {Balagurov}}, \bibinfo {author} {\bibfnamefont {V.}~\bibnamefont {Vaks}},\
  and\ \bibinfo {author} {\bibfnamefont {R.}~\bibnamefont {Zaitsev}},\
  }\bibfield  {title} {\bibinfo {title} {{Statistics of a One-Dimensional Model
  of a Solid Solution}},\ }\href@noop {} {\bibfield  {journal} {\bibinfo
  {journal} {Sov Phys Solid State}\ }\textbf {\bibinfo {volume} {16}},\
  \bibinfo {pages} {1498} (\bibinfo {year} {1975})}\BibitemShut {NoStop}%
\bibitem [{\citenamefont {Shadrin}\ and\ \citenamefont
  {Panov}(2022)}]{Shadrin2022}%
  \BibitemOpen
  \bibfield  {author} {\bibinfo {author} {\bibfnamefont {A.}~\bibnamefont
  {Shadrin}}\ and\ \bibinfo {author} {\bibfnamefont {Y.}~\bibnamefont
  {Panov}},\ }\bibfield  {title} {\bibinfo {title} {Thermodynamic features of
  the {1D} dilute {Ising} model in the external magnetic field},\ }\href
  {https://doi.org/10.1016/j.jmmm.2021.168804} {\bibfield  {journal} {\bibinfo
  {journal} {Journal of Magnetism and Magnetic Materials}\ }\textbf {\bibinfo
  {volume} {546}},\ \bibinfo {pages} {168804} (\bibinfo {year}
  {2022})}\BibitemShut {NoStop}%
\bibitem [{\citenamefont {Panov}(2020)}]{Panov2020}%
  \BibitemOpen
  \bibfield  {author} {\bibinfo {author} {\bibfnamefont {Y.}~\bibnamefont
  {Panov}},\ }\bibfield  {title} {\bibinfo {title} {Local distributions of the
  {1D} dilute {Ising} model},\ }\href
  {https://doi.org/10.1016/j.jmmm.2020.167224} {\bibfield  {journal} {\bibinfo
  {journal} {Journal of Magnetism and Magnetic Materials}\ }\textbf {\bibinfo
  {volume} {514}},\ \bibinfo {pages} {167224} (\bibinfo {year}
  {2020})}\BibitemShut {NoStop}%
\bibitem [{\citenamefont {Panov}\ and\ \citenamefont
  {Rojas}(2021)}]{Panov2021}%
  \BibitemOpen
  \bibfield  {author} {\bibinfo {author} {\bibfnamefont {Y.}~\bibnamefont
  {Panov}}\ and\ \bibinfo {author} {\bibfnamefont {O.}~\bibnamefont {Rojas}},\
  }\bibfield  {title} {\bibinfo {title} {Unconventional low-temperature
  features in the one-dimensional frustrated $q$-state {Potts} model},\ }\href
  {https://doi.org/10.1103/PhysRevE.103.062107} {\bibfield  {journal} {\bibinfo
   {journal} {Physical Review E}\ }\textbf {\bibinfo {volume} {103}},\ \bibinfo
  {pages} {062107} (\bibinfo {year} {2021})}\BibitemShut {NoStop}%
\end{thebibliography}

%
\end{document}